\documentclass{article} 
\usepackage[bbgreekl]{mathbbol}
\let\savedbbchi\bbchi
\usepackage{graphicx}
\usepackage{color}
\usepackage{amsmath}
\usepackage{amsfonts}
\usepackage[normal]{subfigure}
\usepackage{a4wide}
\usepackage{enumitem}
\usepackage{natbib}
\usepackage{authblk}

\renewcommand\harvardyearright[1]{.} 
\let\bbchi\savedbbchi

\newcommand{\cM}{\mathcal{M}}
\newcommand{\scM}{{\scriptscriptstyle \cM}}
\newcommand{\cS}{\mathcal{S}}
\newcommand{\scS}{{\scriptscriptstyle \cS}}

\newcommand{\Ub}{{\bf U}}
\newcommand{\vb}{{\bf v}}
\newcommand{\hv}{\hat{\bf v}}
\newcommand{\scr}{\scriptscriptstyle \rho}
\newcommand{\x}{{\bf x}}
\newcommand{\y}{{\bf y}}

\newcommand{\ie}{{\it i.e.}}
\newcommand{\eg}{{\it e.g.}}
\begin{document}

\title{Modelling physical limits of migration by a kinetic model with non-local sensing}


\author{Nadia Loy \thanks{Department of Mathematical Sciences ``G. L. Lagrange'', Politecnico di Torino, Corso Duca degli Abruzzi 24, 10129 Torino, Italy, and Department of Mathematics ``G. Peano'', Via Carlo Alberto 10 ,10123 Torino, Italy
                (\texttt{nadia.loy@polito.it})}\and
        Luigi Preziosi\thanks{Department of Mathematical Sciences ``G. L. Lagrange'', Dipartimento di Eccellenza 2018-2022, Politecnico di Torino, Corso
                Duca degli Abruzzi 24, 10129 Torino, Italy
                (\texttt{luigi.preziosi@polito.it})}
                \thanks{Corresponding author: \texttt{luigi.preziosi@polito.it}}}
                
\maketitle
\begin{abstract}
Migrating cells choose their preferential direction of motion  in response to different signals and stimuli sensed by spanning their external environment. 
However, the presence of dense fibrous regions, lack of proper substrate, and cell overcrowding  
may hamper cells from moving in certain directions or even from sensing beyond regions that practically act like physical barriers. 
We extend the non-local kinetic model proposed by \cite{Loy_Preziosi} to include situations in which the sensing radius is not constant, but depends on position, sensing direction and time as the behaviour of the cell might be determined on the basis of information collected before reaching physically limiting configurations. 
We analyse how the actual possible sensing of the environment influences the dynamics by recovering the appropriate macroscopic limits and by integrating numerically the kinetic transport equation. 

\end{abstract}

\section{Introduction}
During their motion, cells sense the external environment thanks to their protrusions which may extend up to several cell diameters. The captured chemical or mechanical signals activate transduction pathways inside the cell leading to the cell response which consists in $(i)$ the formation of a ``head" and a ``tail", $(ii)$ in the triggering of actin polimerization at the front edge and depolarization at the rear of the cell, and $(iii)$ in the activation of adhesion molecules and traction forces leading eventually to motion \citep{Aber.53, Adler, Berg_Block_Segall}. The above steps can be somewhat distinguished in polarization and mobility mechanisms. 
For instance, \cite{Devreotes.03} showed that D. Discoideum cells polarize in response to cAMP 
even when treated with inhibitors of the cytoskeleton, such as latrunculin A, that inhibit cell motion. 

After external stimuli  determine the preferential direction of a cell, in addition to internal causes, other environmental cues  may promote or hamper the movement in that direction, such as the cell density. 
For instance, the presence of other cells influences cell migration in a two-fold way. On one hand cells may be attracted due to the mutual interaction of transmembrane adhesion molecules ($\eg$, cadherin complexes). On the other hand, cells may stay away from too crowded regions or just lean on their boundaries. Another important migration determinant is the extra-cellular matrix (ECM), $\ie$ the network of macro-molecules (such as proteoglycans, collagene, fibronectin, and elastin) representing one of the main non-cellular component of all tissues and organs. In fact, it provides cells with a physical scaffold. Its density, stiffness, and microstructure highly influence the behaviour of cells and in particular their migration mode \citep{Boekhorst}. Again, the presence of ECM is necessary to form focal adhesion sites through the activation of integrins that are used by the cell to exert traction forces, but if it is too dense it might represent a steric obstacle to cell motion.

In particular, experiments (see, for instance,  \cite{Schoumacher, Shankar})  show that while the cytoplasm
is very flexible and able to accommodate nearly any pore size (including $1\mu m^2$ gaps
in collagen gels and $0.8\mu m^2$ pores of polycarbonate membranes), the cell nucleus is
five- to ten-fold  stiffer than the surrounding cytoplasm and, with a typical diameter of
$3-10\mu m$, might be larger than the ECM fiber spacing \citep{Davidson}. During
MMP-independent migration ($\ie$, when the proteolytic machinery is inhibited) and in
spite of cell cytoplasm protrusions into the ECM trying to pull the nucleus inside, the
stiff nucleus may be unable to squeeze through narrow pores, setting a critical pore size
below which MMP-inhibited cells remain trapped \citep{Wolf_Friedl.13}. This phenomenon is named physical limit of migration and it has been recently studied from the modelling point of view by \cite{Scianna_Preziosi.13.2, Scianna_Preziosi.13.3, Scianna_Preziosi.14} using cellular Potts models and by \cite{Arduino_Preziosi.15, Giverso.13, Giverso.18} using continuum mechanics methods.

The aim of this article is to include such effects extending the kinetic model developed by \cite{Loy_Preziosi}, where possibly independent cues  are sensed by cells in a non-local way and used to determine respectively cell polarization and speed.
The transport kinetic equation implements then  a velocity jump process that describes the movement performed by the cell as an alternation of runs over straight lines and re-orientations (also called tumbles or turnings) \citep{Stroock}, also considering the bias induced by external stimuli, as done by \cite{Alt.80} and \cite{Alt.88}. Such an equation describes the evolution of a single-particle density distribution like in the Boltzmann equation \citep{Cercignani}. The main elements of the velocity-jump process are the tumbling frequency, the mean speed, and the transition (also called turning or tumbling) probability that describes the probability of choosing a certain velocity after re-orientation. The mean speed, mean runtime (which is the inverse of the frequency), and the tumbling probability may be measured from individual patterns of members of the population.   

In the literature many models considered different sensing strategies and their relation to the determination of cell re-orientation and speed. Focusing on non-local aspects, tipically in position jump processes the transition probabilities depend on the acquisition of information sensed at a certain location ($\eg$ at the target or at the present location) or by averaging the signal over a certain neighborhood as done by \cite{Othmer_Stevens.97} and by \cite{Painter_Sherratt.03}. 

In general, a way of including cell sensing is to consider a non-local average of the external fields. \cite{Othmer_Hillen.02} and \cite{Hillen_Painter_Schmeiser.06} introduced a finite sensing radius and defined a non-local gradient as the average of the external field on a surface which represents the membrane of the cell.  This idea was also used for cell adhesion and haptotaxis by \cite{Armstrong_Painter_Sherratt.06} yielding a macroscopic integro-differential equation. \cite{Butt} recently derived this type of models from a space jump process. Other macroscopic models describing cell migration with non-local measures of the environment were proposed by \cite{Painter_Sherratt.2010, Painter.Gerish.15, Painter_Hillen.02}. \cite{Schmeiser_Nouri} considered a kinetic model with velocity jumps biased towards the chemical concentration gradient.  Similar equations were also proposed in 2D set-ups by \cite{Col_Sci_Tos.15, Col_Sci_Prez.17},  and applied to model crowd dynamics and traffic flow for instance by \cite{Tosin_Frasca}. \cite{Eftimie2} and \cite{Eftimie} also proposed a non-local kinetic models including repulsion, alignement and attraction and used it for modelling tumor dynamics \citep{Eftimie.17} and cell polarisation in heterogeneous cancer cell populations \citep{Bitsouni2018}.

In \cite{Loy_Preziosi} the transition probability of the transport equation  models the probability of choosing a certain velocity direction and speed according to an environmental sensing over a finite neighbourhood of the cell, giving the kinetic model a non-local character. In particular, a double bias is considered, as cells perform a double sensing both of a tactic cue influencing the probability of polarizing in a certain direction and of an environmental cue, usually of mechanical origins, affecting its speed in the polarization direction. For instance, in presence of a chemoattractant giving a preferential direction of motion, a volume filling effect can hamper the motion in a certain direction due to cell overcrowding.

In the present work, then, the focus is on how to include in kinetic models the effect of physical limits of migration.  These include the dependence of the sensing radius on physical characteristics of the environment or on the omotypic or heterotypic cell distributions, that might even hamper the real possibility of cells of measuring in a certain region. In particular, this means that the interval with valuable information to determine cell motion, depends on space, time and direction of sensing.
We also discuss how limiting the sampling volume determines the type of macroscopic limit that can be performed. 

With this aim in mind, the plan of the article is then the following. In Section 2 the kinetic modelling framework is introduced and then specialised to the case of physical limits of migration in Section 3. In particular, the concept of limitation of sensing radius and its effect on motility is qualitatively described. Section 4 focuses on deducing the appropriated macroscopic limit of the kinetic model showing that in presence of physical limits of migration  the macroscopic speed does not vanish in general and therefore the appropriate limit is hyperbolic.
Section 5 starts analysing the model from a simpler case in which there is no cue determining a preferred polarization in space, $\eg$, no chemotaxis, but only mechanical cues influencing cell speed. Simulations of the kinetic model are performed focusing on volume filling effects in Section 5.1 and on cell-ECM interactions in Section 5.2. A particular attention is paid to pointing out what is the effect of different modelling assumptions and how some choices might lead to non-physical results. Then, in Section 6 chemotaxis and cell-cell adhesion are added as mechanisms influencing cell polarization in order to consider the simultaneous presence of physical limits of migration and cell polarization. A final section draws some conclusions pointing out to other open issues.

\section{Modelling framework}\label{sec2}

Let us describe the cell population through the distribution density $p = p(t,\x, v,\hv)$ parametrized by the time $t>0$, the position $\x \in \Omega \subseteq \mathbb{R}^d$, the speed $v\in [0,U]$, where $U$ is the maximal speed a cell can achieve, and the polarization direction $\hv\in \mathbb{S}^{d-1}$ where  $\mathbb{S}^{d-1}$ is the unit sphere boundary in $\mathbb{R}^d$.  
The choice of representing the distribution function $p$ depending on velocity modulus and direction, instead of the velocity vector $\vb=v\hv$, lies in the need of separating the mechanisms governing cell polarization and motility, for instance in response of chemotaxis and in presence of other factors, typically of mechanical origins, influencing cell speed.

The mesoscopic model consists in the transport equation for the cell distribution
\begin{equation}\label{transport.general}
\dfrac{\partial p}{\partial t}(t,\x,v,\hv) + \vb\cdot \nabla p(t,\x,v,\hv) =  \mathcal{J} [p] (t,\x,v,\hv)
\end{equation}
where the operator $\nabla$ denotes the spatial gradient.
The term $\mathcal{J}[p](t,\x,v,\hv)$, named {\it turning operator},
is an  integral operator that describes the change in velocity which is not due to free-particle transport. It may describe the classical run and tumble behaviors, contact guidance phenomena, or cell-cell interactions. For the moment we will consider the classical run and tumble, $\eg$, random re-orientations, which, however, may be biased  by external cues. Therefore, our turning operator will be the implementation of a velocity-jump process in a kinetic transport equation as introduced by \cite{Stroock} and then by \cite{Alt.88}.

Defining $V_p=[0,U]\times \mathbb{S}^{d-1}$, a macroscopic description for the cell population can be classically recovered  through the definition of moments of the distribution function $p$ as follows

\noindent - the cell number density
\begin{equation}\label{def_rho}
\rho(t,\x) = \int_{V_p} p(t,\x,v,\hv) \,dv\,d\hv\,;
\end{equation}

\noindent - the cell mean velocity
\begin{equation}\label{mean.U}
\Ub(t,\x) = \dfrac{1}{\rho(t,\x)}\int_{V_p} p(t,\x,v,\hv)\vb \,dv\,d\hv\,;
\end{equation}

\noindent - the cell variance-covariance matrix
\begin{equation}\label{mean.PP}
\mathbb{P}(t,\x) = \int_{V_p}  p(t,\x,v,\hv) [\vb -\Ub(t,\x)] \otimes [\vb-\Ub(t,\x)] \,dv\,d\hv\,;
\end{equation}

\noindent - the cell speed variance
\begin{equation}\label{mean.E}
E(t,\x) = \int_{V_p} p(t,\x,v,\hv) \frac{|\vb -\Ub(t,\x)|^2}{2}\,dv\,d\hv\,.
\end{equation}

We remark that, because of the definition of $V_p$, the integrals over the velocity space are a simple double integral.
In particular, if the dependence of $f$ on $v$ and $\hv$ can be factorized, $\ie \  f(t,\x,v,\hv)=f^1(t,\x,v) f^2(t,\x,\hv)$, we have that
\begin{equation}
\int_{V_p} f(t,\x,v,\hv) \, dv\,d\hv= \int_{0}^{U} f^1(t,\x,v) \, dv \int_{\mathbb{S}^{d-1}} f^2(t,\x,\hv) \, d\hv,
\end{equation}
where the integral over the boundary of the unit sphere $\mathbb{S}^{d-1}$ is not a surface integral, but  has to be interpreted as
$$\int_{\mathbb{S}^{d-1}} f^2(t,\x,\hv) \, d\hv=\int_{0}^{2\pi} f^2(t,\x,\cos \theta, \sin \theta)\, d\theta\,$$ 
in 2D and similarly in 3D.

The general form of the turning operator which implements a velocity jump processes is
\begin{equation}\label{turning.operator}
\begin{aligned}
\mathcal{J}[p](t,\x,v,\hv) 
& = \int_{V_p} \mu(\x,'\!v,'\!\hv) T(\x,v,\hv|'\!v,'\!\hv)p(t,\x,'\!\!v,'\!\hv)\, d'\!v\,d'\!\hv \\[8pt]
& - \int_{V_p} \mu(\x,v,\hv) T(\x,v',\hv'|v,\hv)p(t,\x,v,\hv)\, dv' d\hv',
\end{aligned}
\end{equation}
where 
$'\!\vb =\, '\!v '\!\hv$ is the pre-turning velocity of the gain term and $\vb'=v'\hv'$ is the post-turning velocity of the loss term.
The so-called {\it turning kernel} $T(\x,v,\hv|'\!v,'\!\hv)$  is the probability for a cell in $\x$ of re-orienting along $\hv$ and moving with speed $v$ given the pre-turning polarization direction $'\!\hv$ and speed $'\!v$.
Being a transition probability, it satisfies 
\begin{equation}\label{normalization.T}
\int_{V_p} T(\x,v',\hv'|v,\hv)\, dv'd\hv' =1 \,,\quad \ \forall \x \, \in \, \Omega,\,\,\forall (v,\hv)\in V_p, 
\end{equation}
which allows to simplify \eqref{turning.operator} to
\begin{equation}\label{turning.operator.2}
\mathcal{J}[p](t,\x,v,\hv)  = \int_{V_p} \mu(\x,'\!\!v,'\!\!\hv) T(\x,v,\hv|'\!v,'\!\!\hv)p(t,\x,'\!\!v,'\!\!\hv)\, d'\!v\,d'\!\hv -\mu(\x,v,\hv) p(t,\x,v,\hv)\,.
\end{equation}

As done by \cite{Stroock, Hillen.05, Chauviere_Hillen_Preziosi.07, Chauviere_Hillen_Preziosi.08}, in the following we will assume that cells retain no memory of their velocity prior to the re-orientation, $\ie$, $T=T(\x,v,\hv)$.
The independence from the pre-tumbling velocity lies in the fact that the choice of the new velocity is linked to the slow interaction process also related to cell ruffling and sensing which is responsible for the biased re-orientation. However, the assumption might be restrictive in some cases, as it excludes, for instance, persistence effects in which the re-orientation direction depends on the pre-tumbling polarization of the cell and the case in which the sensing region depends on the incoming velocity through a polarization-dependent expression of transmembrane receptors.  

Assuming also that the frequency $\mu$ does not depend on the  microscopic velocity allows to simplify considerably 
\eqref{turning.operator.2} in
\begin{equation}\label{J_r.S}
 \mathcal{J}[p](t,\x,v,\hv) = \mu(\x) \, \Big( \rho(t,\x)  T(\x,v,\hv) - p(t,\x,v,\hv) \Big) \,.
\end{equation}
Following \cite{Loy_Preziosi}, we consider here the fact that in taxis processes cells are capable of detecting and measuring external signals  through membrane receptors located along cell protrusions that can extend over a finite radius. Information are then transduced and act as control factors for the dynamics of cells. 
Therefore, in the turning operator of the kinetic model we will include the evaluation of  mean fields  in a neighbourhood of the re-orientation position. In particular, we will consider a control factor $\mathcal{S}$ determining cell polarization (and therefore orientation) and a control factor $\mathcal{S}'$ determining cell speed in that specific orientation direction. 
 Hence, we will have a transition probability that depends on both $\mathcal{S}$ and $\mathcal{S}'$ that can be written as
\begin{equation}\label{distribution.g.r}
T[\cS,\cS'](\x, v,\hv)=c(\x)\int_{\mathbb{R}_+}\gamma_{\scS}(\lambda) \mathcal{T}_{\lambda}^{\hv}[\cS](\x) d\lambda \,  \int_{\mathbb{R}_+}\gamma_{\scS'}(\lambda') \psi(\x,v|\cS'(\x+\lambda'\hv)) \, d\lambda ',
\end{equation}
where $c(\x)$ is a normalization constant, so that according to \eqref{normalization.T} the integral of $T$ over the velocity space is one. In this way, $T$ is a mass preserving transition probability that takes into account two different external fields $\cS$ and $\cS'$ in order to choose the post-tumbling velocity.
Specifically, the quantity $\mathcal{T}_{\lambda}^{\hv}[\cS](\x)$ is a functional which acts on $\cS$ and describes the way the cell measures the quantity $\cS$ around $\x$ along the direction $\hv$ and, therefore, the bias intensity in the direction $\hv$  weighted by the sensing kernel $\gamma_{\scS}(\lambda)$ where $\lambda$ measures the distance from $\x$. In particular,  $\gamma_{\scS}$ has a compact support in $[0,R_{\scS}^{max}]$ where $R_{\scS}^{max}$ is the maximum extension of cell protrusions, $\ie$ the furthest point cells can reach to measure the external signals. In all the applications considered here we take 
\begin{equation}
\mathcal{T}_{\lambda}^{\hv}[\cS](\x) = b\big(\cS(\x + \lambda\hv)\big)\,,
\end{equation}
but other functionals could be considered as discussed by \cite{Loy_Preziosi}.

The density distribution of the modulus 
$\psi=\psi(\x,v|\cS'(\x+\lambda'\hv))$ describes the choice of the speed $v\in[0,U]$ for cells located in $\x$ if oriented along $\hv$ according to the value of another external field $\mathcal{S'}$ measured along this direction and weighted by another function $\gamma_{\scS'}$ with compact support in $[0,R_{\scS'}^{max}]$. 
In fact, the notation $|\cS'$ reads as "given $\cS'$ in a point $\x+\lambda'\hv$ in the neighbourhood of the cell".
In the following the mean of $\psi$, that is normalized to 1 $\ie, \int_{0}^U\psi dv=1$ will be denoted by $\bar{v} (\x|\cS'(\x+\lambda'\hat\vb))$ and its variance by $D (\x|\cS'(\x+\lambda'\hat\vb))$.

\begin{figure}[!h]
\begin{center}\label{Wolf}
 	\includegraphics[width=0.8\textwidth]{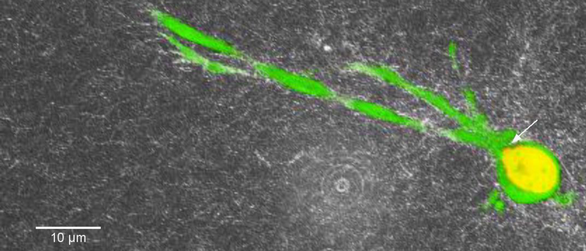}
\end{center}
\caption{Cell extending protrusions toward the  top left corner but unable to move because the nucleus (yellow region on the bottom right corner) can not pass through the pores of the extracellular matrix (by courtesy of P. Friedl and K. Wolf, under permission).}
\end{figure}

\section{Modelling Physical Limits of Migration}

The main novelty of this paper with respect to \cite{Loy_Preziosi} is to consider the existence of physical limits of migration that hamper the cell from sensing or moving beyond a physical barrier. To be specific, for instance, we want to deal with volume filling effects, that occur when cells ahead are too packed for the coming cell to overcome the crowd, or the presence of 
regions with high density of extracellular matrix (ECM), or better, pores so small that the cell nucleus can not pass through them (see, for instance, the work by \cite{Wolf_Friedl.13}).  This means that $R_{\scS}^{max}$ and $R_{\scS'}^{max}$ can not be reached, or even if they are, the decision is taken on the basis of information collected in the space preceding the overcrowded region. Hence, the sensing support of the weight functions $\gamma_{\scS}$ and $\gamma_{\scS'}$ can be reduced.
We mean, for instance, that a cell that is encountering on its way a physical barrier, $\eg$, a basal membrane, might be unable  to squeeze through the fine net, but it can still extend cytoplasmic protrusions beyond the dense area (see Figure \ref{Wolf}). Even if the cell is getting  information, generally denoted by $\cM$, that there would be space to move beyond the barrier, then this does not influence the possibility to pass through it, because the barrier constitutes an impediment to go further. 
However, we observe that in order to test the physical limit of migration we allow the cell to poke a bit in it up to a small depth $\Delta$. For instance, the nucleus in contact with a very small pore in the ECM, still tries to squeeze a bit in, or at least leaning on the entrance of the pore, part of the nucleus will still be in the pore. This also corresponds to the limited, but not absent, freedom
cells have also in constrained situations, like in the case presented in Video 1 of \cite{Wolf_Friedl.13}.

In principle, the limitation on the sensing radius can apply to both the polarization signal $\cS$ and the speed signal $\cS'$, though it is easier to find examples of the latter case. 
In mathematical terms we then assume that there might be a cue $\cM(t,{\bf x})$ related to the sensing of $\cS'$ characterized by a threshold value 
$\cM_{th}$ representing the physical limit of migration. We will sometimes call it a  mechanical signal, though it might be more general than that, like, for instance, the lack of adhesion sites in a certain region to allow cell traction and then migration in a certain area. 
Other examples regard  volume filling when $\cS'$ and $\cM$ are equal to the density of cells $\rho$ and ECM hindrance effects where they are related to the matrix density $M$, or, better, to its characteristic pore cross section.

We then define
 \begin{equation}\label{R_M}
R^{\scM}(t,\x,\hv)=
\begin{cases}
R_{\scS'}^{max} \qquad \textrm{if} \quad \cM(t,\x+\lambda\hv)\le\cM_{th}, \quad \forall \lambda \in[0,R_{\scS'}^{max})\,,\\[10pt]
{\rm inf}\lbrace\lambda\in[0,R_{\scS'}^{max}):\cM(t,\x+\lambda\hat\vb)>\cM_{th} 
\rbrace, \quad {\rm otherwise},  
\end{cases}
\end{equation}
and 
\begin{equation}\label{RMS}
R^{\scM}_{\scS'}(t,\x,\hv)=
{\rm min}\lbrace R^{\scM}(t,\x,\hv)+\Delta, R_{\scS'}^{max}\rbrace\,.
\end{equation}

\begin{figure}[!h]
\begin{center}
 	\includegraphics[scale=0.8]{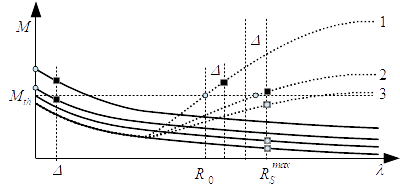}
\end{center}
\caption{
Example of sensing ranges limited by a physical limit of migration corresponding to a level $\cM_{th}$ of $\cM$. Gray circles denote $R^{\scM}$ and black squares $R^{\scM}_{\scS'}$.}
\label{fig:pl1}
\end{figure}

Figure \ref{fig:pl1} highlights the values of $R^{\scM}$ (gray circles) and  $R^{\scM}_{\scS'}$ (black squares) related to several possible landscapes of ${\cal M}$. For instance, the landscape corresponding to the top full line, the cell is already in a region with ${\cM}>{\cM_{th}}$ and so it can hardly move, $R^{\scM}=0$ and  $R^{\scM}_{\scS'}=\Delta$. The same occurs for the second full line from the top. In this case the cell is very close to the border of the barrier, but still inside it. Conversely, the third full line from the top corresponds to a cell at the border of the physical barrier, but that can look freely ahead. In this case,  $R^{\scM}=R^{\scM}_{\scS'}=R^{max}_{\scS}$. Similarly for the bottom full line. However, looking instead at the top dotted line (line 1), the cell encounters a physical barrier again at $\lambda=R_0$, so that $R^{\scM}=R_0$ and  $R^{\scM}_{\scS'}=R_0+\Delta$. For the other two dotted lines (numbered as 2 and 3) the cell is barely touching or not encountering yet the barrier, so that 
$R^{\scM}_{\scS'}=R^{max}_{\scS'}$.

The dependence of $\cM$ on time is due to the fact that the mechanical cue may change in time, for instance because of ECM degradation due to metalloproteinases or redistribution of cell mass in the volume filling case. However, in the following, to simplify a little the notation the dependence on time is usually dropped.

In particular, we remark  that 
\begin{itemize}
\item when the cell center is in a point where $\cM<\cM_{th}$, it means that it  has not reached the physical barrier yet and  $R^{\scM}>0$;
\item when the cell center is in a point where $\cM>\cM_{th}$, then it means that it is stuck in an overcrowded region and $R^{\scM}=0$; 
\item when the cell center is in a point where $\cM=\cM_{th}$, unless for very peculiar cases, it means that it is at the border of the barrier. In this case $R^{\scM}$ vanishes if the cell is polarized towards the barrier and strictly positive if it wants to move away from it.  
\end{itemize}



We also observe that when the presence of a physical barrier limiting $R_{\scS'}^{max}$ to $R^{\scM}_{\scS'}(t,\x,\hv)$ might also limit $R_{\scS}^{max}$ to $R^{\scM}_{\scS}(t,\x,\hv)$.

Summarizing, in order to polarize along the most likely direction $\hv$, the cell with sensing radius $R_{\scS}^{max}$ averages the signal $\cS$ over a region that is described by a weight function $\gamma_{\scS}$ up to  $R_{\scS}^{max}$, or possibly up to a lower value $R^{\scM}_{\scS}(t,\x,\hv)$ if an obstacle is encountered along its sensing activity. Then, the cell polarized along $\hv$ determines its speed averaging another signal $\cS'$ through a weight function $\gamma_{\scS'}$ over the sensing radius $R_{\scS'}^{max}$ or, as above, up to $R^{\scM'}_{\scS'}(t,\x,\hv)$. 

Coming back to Eq.\eqref{distribution.g.r}, as in the work by \cite{Loy_Preziosi}, it is useful to denote the two factors in it as
\begin{equation}\label{B}
B[\cS](\x,\hv)=\int_0^{R^{\scM}_{\scS}(\x,\hv)}
\gamma_{\scS}(\lambda) \mathcal{T}_{\lambda}^{\hv}[\cS](\x)\, d\lambda\,,
\end{equation}
and
\begin{equation}\label{Psi}
\Psi[\cS'](\x,v|\hv)=  \int_0^{R^{\scM'}_{\scS'}(\x,\hv)}
\gamma_{\scS'}(\lambda') \psi(\x,v|\cS'(\x+\lambda'\hv)) \, d\lambda ' \, ,
\end{equation}
in order to highlight the directional and the speed sensing and how they affect the choice of the direction and of the speed independently.
The turning probability in \eqref{J_r.S} then reads
\begin{equation}\label{distribution.g.r.double}
T[\cS,\cS'](\x,v,\vb)=c(\x)B[\cS](\x,\hv)\Psi[\cS'](\x,v|\hv),
\end{equation}
where the normalizing constant is
\begin{equation}\label{c}
c(\x)=\dfrac{1}{\displaystyle \int_{\mathbb{S}^{d-1}} B[\cS](\x,\hv)\Gamma_{\scS'}(\x,\hv) \, d\hv\,},
\end{equation}
with
\begin{equation}
\Gamma_{\scS'}(\x,\hv)=\displaystyle\int_{0}^{R^{\scM'}_{\scS'}(\x,\hv)}
\gamma_{\scS'}(\lambda') \, d\lambda '.
\end{equation}

If $R^{\scM'}_{\scS'}$ is independent of $\hat\vb$, then also $\Gamma_{\scS'}$ is independent of $\hat\vb$ and $c(\x)$ can be factorized  as $c(\x)=c_1(\x)/\Gamma_{\scS'}(\x)$ with 
$$c_1(\x)=\dfrac{1}{\displaystyle \int_{\mathbb{S}^{d-1}} \int_0^{R^{\scM}_{\scS}(\x,\hv)}\gamma_{\scS}(\lambda) \mathcal{T}_{\lambda}^{\hv}[\cS](\x) \,d\lambda \, d\hv},$$
that only depends on the directional sensing.

However, in general, with the definitions \eqref{B} and \eqref{Psi}, the turning operator \eqref{J_r.S} writes
\begin{equation}\label{turning.operator.3}
\mathcal{J}[p](\x,v,\vb) = \mu(\x) \, \Big( \rho(t,\x)c(\x)B[\cS](\x,\hv)
\Psi[\cS'](\x,v|\hv)  - p(t,\x,v,\vb) \Big) \,,
\end{equation}
and the distribution function which nullifies it is
\begin{equation}\label{stationary.eq}
p(t,\x,v,\vb)=\rho(t,\x) c(\x) B[\cS](\x,\hv)\Psi[\cS'](\x,v|\hv).  
\end{equation}
As stated by \cite{Loy_Preziosi}, provided that the probability distribution has initially a finite mass and energy and non-absorbing boundary conditions hold, the function (\ref{stationary.eq}) is a local stable asymptotic equilibrium state \citep{Bisi.Carrillo.Lods, Petterson}. 
The related mean velocity is defined by 
\begin{equation}
{\bf U}_{\scS,\scS'}(\x)=
c(\x) \displaystyle\int_{V_p} B[\cS](\x,\hv)\Psi[\cS'](\x,v|\hv) \vb \,dv\,d\hat\vb \,,
\end{equation}
and may be rewritten as
\begin{equation}
{\bf U}_{\scS,\scS'}(\x)=c(\x)\displaystyle \int_{\mathbb{S}^{d-1}}
\Gamma_{\scS'}(\x,\hv) B[\cS](\x,\hv)\,  
\bar{U}_{\scS'}(\x|\hv) \hv \, d\hv\,,
\end{equation}
where
\begin{equation}\label{barU}
\bar{U}_{\scS'}(\x|\hv)=\dfrac{1}{\Gamma_{\scS'}(\x,\hv)}\displaystyle  \int_0^{R^{\scM'}_{\scS'}(\x,\hv)}
\gamma_{\scS'}(\lambda') \bar{v} (\x|\cS'(\x+\lambda'\hat\vb))\, d\lambda'. 
\end{equation}

Similarly, the variance-covariance tensor of the transition probability 
\begin{equation}
\mathbb{D}_{\scS,\scS'}(\x)=c(\x) \displaystyle \int_{V_p}
B[\cS](\x,\hv) \,  
\Psi[\cS'](\x,v|\hv)
 (\vb-{\bf U}_{\scS,\scS'}(\x)) \otimes 
(\vb-{\bf U}_{\scS,\scS'}(\x))\, \,dv\,d\hat\vb\,,
\end{equation}
can be written as 
\begin{equation}
\mathbb{D}_{\scS,\scS'}(\x)=c(\x) \displaystyle \int_{\mathbb{S}^{d-1}}
\Gamma_{\scS'}(\x,\hv) B[\cS](\x,\hv)\,  
\bar{D}_{\scS'}(\x|\hv)
 \hv\otimes\hv \, d\hv-
{\bf U}_{\scS,\scS'}(\x) \otimes {\bf U}_{\scS,\scS'}(\x)\,,
\end{equation}
where
\begin{equation}\label{barD}
\bar{D}_{\scS'}(\x|\hv)=
\dfrac{1}{\Gamma_{\scS'}(\x,\hv)}\displaystyle  \int_0^{R^{\scM'}_{\scS'}(\x,\hv)}
\gamma_{\scS'}(\lambda') D (\x|\cS'(\x+\lambda'\hat\vb))\, d\lambda'. 
\end{equation}

In order to understand the meaning of $\bar{U}_{\scS'}$ in the presence of physical limits of migration, let us consider the following case: assume that ahead of a cell there is a space free of obstacles up to a distance $R_0$, $\eg$, the ECM density has a density yielding a mean speed $\bar v_0$, while after $R_0$ the ECM density is so large to represent a physical limit of migration with a much lower mean speed, say $\bar{\varepsilon}$ or even vanishing with $\bar{v}=0$ and $\psi=\delta(v)$. For sake of simplicity, let us assume that $\gamma_{\scS'}$ is a Heaviside function.  
The following interesting cases are possible
\begin{itemize}

\item If $R_0>R_{\scS'}^{max}$, then the critical value for   $\cM$ is located beyond the sensing radius (as for the dashed line 1 in Fig. \ref{fig:pl1}), so that
$R^{\scM}_{\scS'} = R^{\scM}= R_{\scS'}^{max}$, and finally
$\bar{U}_{\scS'}=\bar v_0$;

\item If $R_{\scS'}^{max}-\Delta<R_0<R_{\scS'}^{max}$ (as for the dashed line 2 in Fig. \ref{fig:pl1}), then the critical value for   $\cM$ is barely sensed with the cell poking a bit in the region with $\cM>\cM_{th}$. In this case
$R^{\scM}=R_0$ but 
$R^{\scM}_{\scS'} =R_{\scS'}^{max}$, so that 
$\bar{U}_{\scS'}=\dfrac{\bar v_0 R_0+\bar{\varepsilon}(R_{\scS'}^{max}-R_0)}{R_{\scS'}^{max}}$;

\item If $R_0<R_{\scS'}^{max}-\Delta$ (as for the dashed line 3 in Fig. \ref{fig:pl1}), then the cell pokes for a depth 
$\Delta$ into the region with $\cM>\cM_{th}$. In this case
$R^{\scM}=R_0$ and 
$R^{\scM}_{\scS'} =R_0+\Delta$, so that 
$\bar{U}_{\scS'}=\dfrac{\bar v_0 R_0+\bar{\varepsilon}\Delta}{R_0+\Delta}$;

\item In particular, if $R_0=0$,  the cell has arrived at the barrier, then
$R^{\scM}=0$ and 
$R^{\scM}_{\scS'} =\Delta$, so that 
 $\bar{U}_{\scS'}=\bar{\varepsilon}$.
Hence, in the limit $\Delta\to 0$ when the cell is approaching the physical limit, its speed decreases fast to $\bar \varepsilon$, which can also be taken to be zero, as in the simulations to follow;

\item Conversely, if the cell is so close to an exit, say at a distance $R_{exit}\in[0,\Delta]$, that it can sense outside the physical barrier, then $R_0=0$ 
and 
$R^{\scM}_{\scS'} =\Delta$, so that 
$\bar{U}_{\scS'}=\bar v_0-(\bar v_0-\bar \varepsilon)\dfrac{R_{exit}}{\Delta}$
Hence, in the limit $\Delta\to 0$ its speed increases fast to $\bar v_0$.

\end{itemize}

\begin{figure}[!h]
\begin{center}
 	\includegraphics[width=0.6\textwidth]{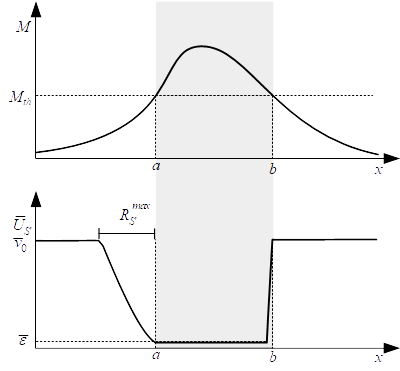}
\end{center}

\caption{Speed of a cell  polarized to move to the right according to its position in presence of a speed limit region in the gray zone $(a,b)$, for instance due to too dense ECM to allow cell migration. }
\label{fig:pl2}
\end{figure}

Figure \ref{fig:pl2} gives an example of the measured $\bar{U}_{\scS'}$ as a function of the cell position (with the cell polarized to move to the right) when the region with  $\cM >\cM_{th}$ is the interval $(a,b)$. The cell moves at a velocity $\bar v_0$ till it reaches a distance $R^{max}_{\scS'}$ from the barrier. Then it decreases its speed at first only sligthly and then faster when getting closer to the barrier to reach a very small or vanishing value. If on the other hand the cell is at the right border of the barrier it can move freely with velocity $\bar v_0$.

In the simulations to follow, in order to test the physical limit of migration, we shall always consider $\Delta=\bar{\varepsilon}=0$, $\ie$, the case in which the cell can not poke in the limited area. We leave to a future work the investigation of the effect of considering a non-vanishing $\Delta$.

\section{Macroscopic limits}
One of the main issues about transport equations consists in recovering the appropriate macroscopic limit allowing to highlight the driving macroscopic phenomenon. 
Typically these are obtained by identifying a small parameter $\epsilon\ll 1$ that allows to suitably rescale the transport equation with a parabolic or a hyperbolic scaling.  These correspond, respectively, to 
$$
\begin{array}{lr}
\tau=\epsilon^2 t, \quad {\bf \boldsymbol{\xi}} =\epsilon \x, \\[8pt]
\tau=\epsilon t, \quad {\bf \boldsymbol{\xi}} =\epsilon \x.
\end{array}
$$
The proper choice of the time scale comes from a nondimensionalization of the turning operator \eqref{J_r.S}. 
Referring to the functions introduced in (\ref{B}) and (\ref{Psi}), we suppose that, up to a nondimensionalization, $B[\cS]$ and $\Psi[\cS']$ may be written as
$$B[\cS]=B[\cS]_0+\epsilon B[\cS]_1+\mathcal{O}(\epsilon^2),$$
$$\Psi[\cS']=\Psi[\cS']_0+\epsilon\Psi[\cS']_1 +\mathcal{O}(\epsilon^2),$$  
which describe different orders of bias. Then, the turning kernel $T[\cS,\cS']$ can be written as
\begin{equation}\label{T.expanded}
T[\cS,\cS'](\boldsymbol{\xi}, v,\hat{\vb})=T[\cS,\cS']_0(\boldsymbol{\xi}, v,\hat{\vb})+\epsilon T[\cS,\cS']_1(\boldsymbol{\xi}, v,\hat{\vb})+\mathcal{O}(\epsilon^2), 
\end{equation}
where
\begin{equation}\label{T0}
T[\cS,\cS']_0(v,\hat{\vb})=c(\x)B[\cS]_0(\hv)\Psi[\cS']_0(v|\hv),
\end{equation}
and
\begin{equation}\label{T1}
T[\cS,\cS']_1(v,\hat{\vb})=c(\x)B[\cS]_0(\hv)\Psi[\cS']_1(v|\hv)+c(\x)B[\cS]_1(\hv)\Psi[\cS']_0(v|\hv)\,.
\end{equation}
Coherently, the means and variances of $\Psi[\cS']_0$ and $\Psi[\cS']_1$ will be, respectively, denoted by $\bar U_{\scS'}^0$, $\bar U_{\scS'}^1$, $\bar{D}_{\scS'}^0$, $\bar{D}_{\scS'}^1$. 
Similarly, the distribution function $p$ is expanded as
\begin{equation}
p=p_0+\epsilon p_1 +\mathcal{O}(\epsilon^2).
\end{equation}
 As there is conservation of mass we have that  \citep{Othmer_Hillen.00}
all the mass is in $p_0$, $\ie$,
\begin{equation}\label{rho0}
\rho_0=\rho, \quad \rho_i=0 \quad \forall   i \geq 1 \, ,
\end{equation}
where $\displaystyle{\rho_i=\int_{V_p}p_i \, dv\, d\hat{\vb}}$. 
Furthermore, for performing the diffusive limit we shall suppose that $\displaystyle{\int_{V_p} p_i v\hat{\vb}\, dv\, d\hat{\vb} =0 } \quad \forall i \geq 2$. 

If we suppose that
$$\displaystyle \int_{0}^U \Psi[\cS']_0(v|\hv)\, dv=\displaystyle \int_{0}^U \Psi[\cS'](v|\hv)\, dv\,, \qquad 
\displaystyle \int_{0}^U \Psi[\cS']_i(v|\hv)\, dv=0 \quad \forall i\geq 1,$$
and
$$\displaystyle \int_{\mathbb{S}^{d-1}} B[\cS]_0(\hv) \, d\hv=\int_{\mathbb{S}^{d-1}} B[\cS](\hv) \, d\hv\,,\qquad
\int_{\mathbb{S}^{d-1}} B[\cS]_i(\hv) \, d\hv=0\quad \forall i\geq 1,$$
then 
\begin{subequations}
\begin{equation}\label{diff.cond.1}
\displaystyle \int_{V_p} T[\cS,\cS']_0(\boldsymbol{\xi}, v,\hat{\vb})  \, dv\, d\hat{\vb} =1,
\end{equation}
and
\begin{equation}\label{diff.cond.2}
\displaystyle \int_{V_p} T[\cS,\cS']_i(\boldsymbol{\xi}, v,\hat{\vb})  \, dv\, d\hat{\vb} =0 \quad \forall i \geq 1,
\end{equation}
\end{subequations}
that are needed for performing a macroscopic limit \citep{Othmer_Hillen.00}.

Denoting by $\mathcal{J}^i$ the turning operators defined by $T[\cS,\cS']_i$, it turns out for instance that
\begin{equation}\label{Jr0}
\mathcal{J}^0[p_0](v,\hat{\vb})=\mu(\rho_0T[\cS,\cS']_0(v,\hat{\vb})-p_0(v,\hat{\vb}))=0 \, ,
\end{equation}
and therefore 
the function $p_0$ is the equilibrium function given by 
\begin{equation}\label{p0}
p_0(\vb_p)=\rho_0 T[\cS,\cS']_0 (v,\hat{\vb}).
\end{equation}

The velocity and tensor of the second moments of the transition probability then naturally write as
\begin{equation}
\Ub_{\scriptscriptstyle\cS,\mathcal{S'}}=\Ub_{\scriptscriptstyle\cS,\mathcal{S'}}^0+\epsilon \Ub_{\scriptscriptstyle\cS,\mathcal{S'}}^1 +\mathcal{O}(\epsilon^2)\,,
\end{equation} 
and
\begin{equation}
\mathbb{D}_{\scriptscriptstyle\cS,\mathcal{S'}}=\mathbb{D}_{\scriptscriptstyle\cS,\mathcal{S'}}^0+\epsilon \mathbb{D}_{\scriptscriptstyle\cS,\mathcal{S'}}^1+\mathcal{O}(\epsilon^2)\,,
\end{equation}
where 
$$\Ub_{\scriptscriptstyle\cS,\mathcal{S'}}^i=\displaystyle\int_{V_p}T[\cS,\cS']_i \vb \, dv\, d\hat{\vb}\,,$$ 
and 
$$\mathbb{D}_{\scriptscriptstyle\cS,\mathcal{S'}}^i=\displaystyle\int_{V_p} T[\cS,\cS']_i (\vb-\Ub_{\scriptscriptstyle\cS,\mathcal{S'}}^i) \otimes (\vb-\Ub_{\scriptscriptstyle\cS,\mathcal{S'}}^i) \, dv\, d\hat{\vb}\,,$$ 
for $i\geq 0$.
Then, using \eqref{T0} and \eqref{T1}, the macroscopic velocity of order 0 is
\begin{equation}\label{macro.drift.double}
{\bf U}_{\scS,\scS'}^0(\boldsymbol{\xi})=
c(\boldsymbol{\xi})\int_{\mathbb{S}^{d-1}}
\Gamma_{\scS'}(\boldsymbol{\xi},\hv) B[\cS]_0(\hv) \,  
\bar{U}_{\scS'}^0(\boldsymbol{\xi}|\hv)
\hv \, d\hv\, ,
\end{equation}
that of order 1 is
\begin{equation}
{\bf U}_{\scS,\scS'}^1(\boldsymbol{\xi})=
c(\boldsymbol{\xi})\int_{\mathbb{S}^{d-1}} 
\Gamma_{\scS'}(\boldsymbol{\xi},\hv)
\Big(
B[\cS]_0(\hv) \bar{U}_{\scS'}^1(\boldsymbol{\xi}|\hv)
+B[\cS]_1(\hv)\bar{U}_{\scS'}^0(\boldsymbol{\xi}|\hv)
\Big)\hv \, d\hv
\end{equation}
and the equilibrium diffusion tensor is
\begin{equation}
\mathbb{D}_{\scS,\scS'}(\x)=c(\boldsymbol{\xi}) \displaystyle \int_{\mathbb{S}^{d-1}}
\Gamma_{\scS'}(\boldsymbol{\xi},\hv) B[\cS]_0(\boldsymbol{\xi},\hv)\,  
\bar{D}_{\scS'}^0(\boldsymbol{\xi}|\hv) \hv\otimes\hv \, d\hv-
{\bf U}_{\scS,\scS'}^0(\boldsymbol{\xi}) \otimes {\bf U}_{\scS,\scS'}^0(\boldsymbol{\xi})\,,
\end{equation}
where
$$\bar{D}_{\scS'}^0(\boldsymbol{\xi}|\hv)=\dfrac{1}{\Gamma_{\scS'}(\boldsymbol{\xi},\hv)}
\displaystyle \int_0^{R^{\cM'}_{\scS'}(\boldsymbol{\xi},\hv)}
\gamma_{\scS'}(\lambda') D^0(\boldsymbol{\xi}|\cS'(\boldsymbol{\xi}+\lambda'\hat\vb))\, d\lambda' .$$ 
The diffusion tensor is in general anisotropic, as $\bar{D}_{\scS'}^0$  may be non isotropic  in $\hv$.

The discriminating factor on which scaling can be performed is whether $T$ is such that
\begin{equation}\label{even}
\Ub_{\scS,\scS'}^0 = \textbf{0}
\end{equation}
$\ie$, the leading order macroscopic velocity vanishes, or not. This is the necessary functional condition for performing a diffusive limit, and it is actually what we expect in a diffusive regime.
Referring to Eq. \eqref{macro.drift.double}, we observe that, even in the case in which $B[\cS]$ is constant, corresponding for instance to no directional bias and $R_{\scS'}^{\scM'}$, and therefore $\Gamma_{\scS'}$ not depending on $\hat{\vb}$, Eq. \eqref{even} is satisfied if the mean velocity 
$\bar{U}_{\scS'}^0(\boldsymbol{\xi}|\hv)$ is even as a function of $\hat{\vb}$ for $a.e.\ \boldsymbol{\xi}\in \Omega$. Moreover, if $R_{\scS'}^{\scM'}$ depends on $\hat{\vb}$, condition \eqref{even} can not be satisfied, unless for trivial cases that do not include physical limits of migration, as we will illustrate in the next sections through some examples. 
Therefore, in general one can only perform a hyperbolic limit which reads
\begin{equation}\label{macro.drift}
\dfrac{\partial \rho}{\partial \tau}+\nabla \cdot \big( \rho \Ub_{\scS,\scS'}^0 \big)=0 \, .
\end{equation}

In order to see which regimes are diffusive or hyperbolic, it is instructive to see what happens if $R_{\scS'}(\x,\hv)$ is much smaller then the characteristic length of variation of $\cS'$ as done by \cite{Loy_Preziosi} for the directional bias. In this case, we can expand $\cS'$ as
\begin{equation}\label{S_approx}
\cS'(\x+\lambda'\hat{\vb})=\cS'(\x)+\lambda'\hv\cdot\nabla\cS'(\x) +\mathcal{O}(\lambda'^2) \quad \forall \lambda' \le R_{\scS'}(\x,\hv)
\end{equation}
and we have that that the quantity $\cS'(\x)+\lambda'\hv\cdot\nabla\cS'(\x)$ stays positive.
The density function $\psi(v|\cS'(\y))$, $\y \in \Omega$ may be seen as a function of two variables
\[
\tilde{\psi}: (v,\y)\in[0,U]\times\Omega \longmapsto \tilde{\psi}(v,\y)=\psi(v|\cS'(\y))
\]
composed with the function $\cS'$ defined on $\Omega$.
If we suppose that $\tilde{\psi} \in L^1([0,U])\times C^2(\Omega)$
with
\begin{equation}\label{psi_norm}
\int_{0}^U\tilde{\psi}(v,\y)dv=\int_0^U\psi(v|\cS'(\y))dv=1,
\end{equation}
then, in virtue of \eqref{S_approx} we may write
\begin{eqnarray*}
\tilde{\psi}(v,\cS'(\x+\lambda'\hv))&=&
\tilde{\psi}(v,\cS'(\x)+\lambda'\hv\cdot\nabla\cS'(\x)+\mathcal{O}(\lambda'^2)) \\
&=& \tilde{\psi}(v,\cS'(\x))+\dfrac{\partial}{\partial \cS'}[\tilde{\psi}(v,\cS'(\x))]\lambda'\hv\cdot\nabla\cS'(\x)+
\mathcal{O}(\lambda'^2) \quad\forall\lambda'\le R_{\scS'}(\x,\hv).
\end{eqnarray*}
Therefore, recalling Eqs.\eqref{distribution.g.r.double} and \eqref{c} the transition probability writes
\begin{eqnarray*}
T(\x,v,\hv)&=&
\dfrac{B[\cS](\hv)}{\displaystyle{\int_{\mathbb{S}^{d-1}}B[\cS](\hv)\Gamma_{\scS'}(\x,\hat{\vb})\,d\hv}}
\int_0^{R_{\scS'}(\x,\hv)}\gamma_{\scS'}(\lambda')\psi(v|\cS'(\x+\lambda'\hv))\,d\lambda'\\
&\approx &
\dfrac{B[\cS](\hv)\Gamma_{\scS'}(\x,\hat{\vb})}{\displaystyle{\int_{\mathbb{S}^{d-1}}B[\cS](\hv)\Gamma_{\scS'}(\x,\hat{\vb})\,d\hv}}
\left[\psi(v|\cS'(\x))+\Lambda'(\x,\hv)\dfrac{\partial}{\partial \cS'}[\psi(v|\cS'(\x))] \nabla\cS'(\x)\cdot\hv\right] 
\end{eqnarray*}
where
$$\Lambda'(\x,\hat{\vb})=\dfrac{1}{\Gamma_{\scS'}(\x,\hat{\vb})}
\int_0^{R_{\scS'}(\x,\hv)}\gamma_{\scS'}(\lambda')\lambda'\,d\lambda'.$$
The quantity
\[
\dfrac{\partial}{\partial \cS'}\bar{v}(\x|\cS'(\x))
\]
is the derivative of the master curve relating $\bar v$ to the signal $\cS'$.

Considering the rescaling ${\bf \boldsymbol{\xi}} =\epsilon \x$, one has that 
$$T[\cS,\cS']_0=
\dfrac{B[\cS](\hv)\Gamma_{\scS'}({\bf \boldsymbol{\xi}},\hat{\vb})}{\displaystyle{\int_{\mathbb{S}^{d-1}}B[\cS](\hv)\Gamma_{\scS'}({\bf \boldsymbol{\xi}},\hat{\vb})\,d\hv}}
\psi(v|\cS'({\bf \boldsymbol{\xi}}))
$$
and 
$$T[\cS,\cS']_1=
\dfrac{B[\cS](\hv)\Gamma_{\scS'}({\bf \boldsymbol{\xi}},\hat{\vb})\Lambda'({\bf \boldsymbol{\xi}},\hat{\vb})}
{\displaystyle{\int_{\mathbb{S}^{d-1}}B[\cS](\hv)\Gamma_{\scS'}({\bf \boldsymbol{\xi}},\hat{\vb})\,d\hv}}
\dfrac{\partial}{\partial \cS'}[\tilde{\psi}(v,\cS'({\bf \boldsymbol{\xi}}))] \nabla\cS'({\bf \boldsymbol{\xi}})\cdot\hv\,.
$$
Integrating over the velocity space $V_p$ it can be readily checked that $T_0$ satisfies condition \eqref{diff.cond.1} and $T_1$ 
satisfies condition \eqref{diff.cond.2} observing that 
\[
\int_0^U\dfrac{\partial}{\partial \cS'}\psi(v|\cS'({\bf \boldsymbol{\xi}}))\,dv=\dfrac{\partial}{\partial \cS'}\int_0^U\psi(v|\cS'({\bf \boldsymbol{\xi}}))\,dv =0.
\]
One then has that 
\begin{equation}\label{U_macro_gen}
\Ub_{\scS'}^0= \bar{v}({\bf \boldsymbol{\xi}}|\cS'({\bf \boldsymbol{\xi}})){\bf N}_{\scS'}\,,\quad{\rm where}\quad
{\bf N}_{\scS'}=\dfrac
{\displaystyle{\int_{\mathbb{S}^{d-1}}B[\cS](\hv)\Gamma_{\scS'}(\hv)\hv\, d\hv}}
{\displaystyle{\int_{\mathbb{S}^{d-1}}B[\cS](\hv)\Gamma_{\scS'}(\hv)     \, d\hv}}\,, 
\end{equation}
and
$$\Ub_{\scS'}^1=\dfrac{\partial}{\partial \cS'}[\bar{v}({\boldsymbol{\xi}}|\cS'({\boldsymbol{\xi}}))]\mathbb{T}_{\scS'}\nabla \cS'
\,,\quad{\rm where}\quad
\mathbb{T}_{\scS'}=\dfrac
{\displaystyle{\int_{\mathbb{S}^{d-1}}B[\cS](\hv)\Gamma_{\scS'}(\hv)\hv\otimes\hv\, d\hv}}
{\displaystyle{\int_{\mathbb{S}^{d-1}}B[\cS](\hv)\Gamma_{\scS'}(\hv)     \, d\hv}}\,. 
$$
We then observe that the mean direction ${\bf N}_{\scS'}$  defined in Eq. \eqref{U_macro_gen} of the cell population may hardly be zero, except in the case in which $B$ and $\Gamma_{\scS'}$ are even in every direction, $\ie$
\begin{equation}\label{pari}
B[\cS](\boldsymbol{\xi},\hv)=B[\cS](\boldsymbol{\xi},-\hv) \quad {\rm and} \quad \Gamma_{\scS'}(\boldsymbol{\xi},\hv)=\Gamma_{\scS'}(\boldsymbol{\xi},-\hv) \quad a.e. \quad {\rm in } \quad \Omega.
\end{equation}
This may happen only if $R_{\scS'}$ is even as a function of the direction $\hv$, $\eg$, if it is constant which means that there is no physical limit of migration. As soon as a physical barrier appears, in points close to it an asymmetry appears in the evaluation of the sensing radius which make \eqref{pari} not valid.

In the simulations to follow, we shall illustrate that because of these reasons, even if $R_{\scS}^{max}$ is small as well or in the case of isotropic polarization $B=const$, the mean velocity hardly vanishes close to physical barriers and, then, Eq. \eqref{even} is not satisfied, leading to the necessity of performing a hyperbolic scaling with a parabolic scaling only possible away from the barriers.

For sake of completeness, we recall that in these cases the diffusive limit would lead to
\begin{equation}\label{macro.diff}
\dfrac{\partial \rho}{\partial \tau}+\nabla \cdot \Big(\rho \Ub_{\scS,\scS'}^1 \Big)=\nabla \cdot \left( \dfrac{1}{\mu}\nabla \cdot \big(\mathbb{D}_{\scS,\scS'}^0\rho \big)\right)\, .
\end{equation}

\subsection{Boundary conditions}
We shall consider the case in which there is conservation of mass in the domain of integration. Therefore, we will apply 
biological no-flux condition \citep{Plaza}
\begin{equation}\label{noflux}
\int_{V_p} p(\tau,\boldsymbol{\xi},v,\hv) \hv\cdot {\bf n}(\boldsymbol{\xi})\,dv\,d\hat\vb=0, \quad \forall \boldsymbol{\xi} \in \partial \Omega, \quad \tau>0
\end{equation}
being ${\bf n}(\boldsymbol{\xi})$ the outer normal to the boundary $\partial \Omega$ in the point $\boldsymbol{\xi}$. This class of boundary conditions is part of the wider class of non-absorbing boundary conditions. 
At the macroscopic level \eqref{noflux} gives \citep{Plaza}
\[
\Big(  \mathbb{D}_{\scriptscriptstyle\mathcal{S},\mathcal{S}'}\nabla\rho -\rho \Ub_{\scriptscriptstyle\mathcal{S},\mathcal{S}'}^1\Big)\cdot {\bf n}=0, \quad {\rm on} \quad \partial \Omega,
\]
for the diffusive limit, whilst for the hyperbolic limit the corresponding boundary condition is 
\[
\Ub^0\cdot {\bf n}=0, \quad \rm{on} \quad \partial \Omega.
\]

There are two important classes of kinetic boundary conditions which satisfy \eqref{noflux}: the regular
reflection boundary operators and the non-local (in velocity) boundary operators of diffusive type. We address the reader to the works by \cite{Palc} and by \cite{Lods} for the definition of these boundary operators. In the present work, we shall consider  Maxwell-type boundary conditions which are prescribed in the form
\begin{equation}\label{Maxwell}
p(\tau,\boldsymbol{\xi}, v',\hat{\vb}')=
\alpha(\boldsymbol{\xi})p(\tau,\boldsymbol{\xi}, v,\mathcal{V}(\hat{\vb}))+
(1-\alpha(\boldsymbol{\xi}))M(\boldsymbol{\xi},v, \hat\vb)\int_{ \hat{\vb}^*\cdot {\bf n}\ge 0} p(\tau,\boldsymbol{\xi},v^*,\hat{\vb}^*) |\hat \vb^*\cdot {\bf n}|\,dv^*\,d\hv^*,
\end{equation} 
where $\mathcal{V}(\hat{\vb})=-\hat{\vb}$ for the bounce back reflection condition and $\mathcal{V}(\hat{\vb})=\hat{\vb}-2(\hat{\vb}\cdot{\bf n}){\bf n}$ for the specular reflection. $M(\boldsymbol{\xi},v,\hat\vb)$ is the Maxwellian function at the wall of $\Omega$.

\subsection{Numerical aspects}\label{Numerics}
The numerical scheme is the same used by \cite{Loy_Preziosi}.
We consider a computational domain in the form $\Omega\times V_p$ where in the one dimensional case 
$\Omega=[x_{min},x_{max}]$ and $V_p$ symmetrized considering the two only possible directions along $x$ and $-x$ in $\tilde V_p=[-U,U]$ and in the two dimensional case 
$\Omega=[x_{min},x_{max}]\times[y_{min},y_{max}]$ and $V_p=[0,U]\times[0,2\pi]$.
The computational domain is discretized with a Cartesian mesh ${\bf X}_h\times {\bf V}_{p_h}$, where ${\bf X}_h$ and ${\bf V}_{p_h}$ are defined by (in two dimensions)
\begin{eqnarray*}
{\bf X}_{h} &=&\lbrace \x_{ij}=(x_i,y_j)=(x_{min}+i\Delta x, y_{min}+j\Delta y), \quad i=0,\ldots,n_x, \quad  j=0,\ldots,n_y \rbrace \\[20pt]
{\bf V}_{p_h} &=&\lbrace \vb_{l,k}=v_k(\cos \theta_l,\sin \theta_l),\quad \theta_l=(j+1/2)\Delta \theta, 
\quad l=0,\ldots, n_{ang}-1, v_k=v_0+k\Delta v,\quad k=0,\ldots,n_v \rbrace
\end{eqnarray*}
where $\Delta x=\dfrac{x_{max}-x_{min}}{n_y}$, $\Delta y=\dfrac{y_{max}-y_{min}}{n_y}$, $\Delta v=\dfrac{U}{n_v}$, $\Delta \theta =\dfrac{2\pi}{n_{ang}}$.
Denoting by $p^n_{i,j,l,k}$ an approximation of the distribution function $p(t^n,\x_{i,j},\vb_{p_{l,k}})$, where $\vb_{p_{l,k}}=(v_k,\hat{\vb}_l)$.
We introduce the first order splitting
\[
\begin{cases}
\dfrac{p_{i,j,l,k}^{n+1/2}-p_{i,j,l,k}^n}{\Delta t}+\vb \cdot \nabla_{\x,h}p_{i,j,l,k}^n=0 \qquad \rm{(transport \ step)}\\[20pt]
\dfrac{p_{i,j,l,k}^{n+1}-p_{i,j,l,k}^{n+1/2}}{\Delta t}=\mu\Big(\rho_{i,j}^{n+1}T[\mathcal{S},\mathcal{S}']_{i,j,l,k}-p_{i,j,l,k}^{n+1}\Big) \qquad \rm{(relaxation \ step)}
\end{cases}
\]
where $h=(\Delta t, \Delta x, \Delta y)$, $\vb \cdot \nabla_{\x,h}p_{i,j,l,k}^n$ is an approximation of the transport operator $\vb \cdot \nabla p$ computed with a Van Leer scheme. It is a high resolution monotone, conservative scheme which is second order if the solution is smooth and first order near the shocks. $T[\mathcal{S},\mathcal{S}']_{i,j,k}$ is the discretization of the transition probability $T$. We observe that as the turning operator preserves mass and the turning probability is known and does not depend on $p$, the relaxation step may be implicit and we may consider the density at time $n+1$. In particular the density is computed by using a trapezoidal rule
\[
\rho_{i,j}^n=\Delta v \Delta \theta\sum_{k=0}^{n_v}\sum_{j=0}^{n_{ang}}p_{i,j,l,k}^n .
\]

Concerning boundary conditions, in the one-dimensional case we consider regular reflecting conditions. In one dimension, the bounce-back and the specular reflection boundary conditions coincide, that is $p(t,x=x_{min},v)=p(t,x=x_{min},-v)$ and $p(t,x=x_{max},-v)=p(t,x=x_{max},v)$. We do not consider Maxwell type conditions as only the outgoing speed would be affected.
In the two-dimensional case, the regular reflection is biologically unrealistic, as cells do not bounce back nor they collide with the wall as hard spheres. Therefore, Maxwell type boundary conditions are more realistic, and we shall consider for the Maxwellian to the wall
\[
M(\x,v,\hv)=T[\mathcal{S},\mathcal{S}'](\x,v,\hv),
\]
being $T$ the asymptotic equilibrium of the system with this class (no-flux) of boundary conditions. 

Concerning the relaxation step, we remark that $T[\cS,\cS']$ is defined as in Eq. \eqref{distribution.g.r.double} where $\Psi$ is a probability density with a minimal variance, as numerically we can not represent a Dirac delta, even because this would require a weak formulation of all the equations. We then used
\[
\Psi[\cS'](\x,v|\hv)=C\exp\left[ -\,k\cos \left(\dfrac{2\pi(v-\bar{U}_{\scS'})}{U} \right)\right], \qquad v\in[0,U]
\] 
that is a Von Mises distribution translated on the speed interval $[0,U]$ with $C$ a normalization constant. Its average is exactly $\bar{U}_{\scS'}$ that we compute using the definition \eqref{barU}. In order to approximate a Dirac delta we shall consider small variances of $\Psi$ and, then, large values of $k$. Numerically, though, the variance of $\Psi$ will be larger then $\Delta v$ being $\Delta v$ the minimum distance between two possible values of the speed, because if it is smaller it causes numerical spurious errors.

\section{Random polarization}\label{isotropicpol}

The environmental cues which may represent physical limits of migration in general affect cell speed. Therefore, we here first focus on the particular case in which there is no bias in the decision of the direction of motion, corresponding to a random polarization, given by $B[\cS](\x,\hat\vb)=const$. As said by \cite{Loy_Preziosi}, this does not imply isotropy, because the speed can have different density distributions on every direction $\hat{\vb}$, as the distribution of $\cS'$ sensed ahead along the direction $\hat{\vb}$ may be different. Furthermore, in this article, such differences of $\cS'$ may also lead to different sensing radii in different directions $\hv$.  

In this section we will specify the model in the random polarization case and, eventually, we will introduce some practical examples characterized by anisotropy on the sensing radius depending on the direction, as well as on space and time.

If  in Eq.\eqref{B} $\mathcal{T}_{\lambda}^{\hv}[\cS]$ is independent of $\hv$, then the transition probability
 (\ref{distribution.g.r.double}) simplifies to 
\begin{equation}\label{TSR0half}
T[\cS,\cS'](\x,v,\hat{\vb})=\dfrac{1}{\displaystyle\int_{\mathbb{S}^{d-1}}\Gamma_{\scS'}(\x,\hv)d\hv}
\int_0^{R_{\scS'}^{\scM'}(\x,\hv)}\gamma_{\scS'}(\lambda') \psi(\x,v|\cS'(\x+\lambda'\hv)) \, d\lambda' \,,
\end{equation}
where we recall that 
\[
\Gamma_{\scS'}(\x,\hat{\vb})=\int_0^{R_{\scS'}^{\scM'}(\x,\hv)}\gamma_{\cS'}(\lambda')\,d\lambda'.
\]
The macroscopic velocity of the transition probability simplifies to
\begin{equation}\label{U0.rp}
\Ub_{\scS'}(\x)=\dfrac{1}{\displaystyle\int_{\mathbb{S}^{d-1}}\Gamma_{\scS'}(\x,\hv)d\hv} 
\displaystyle \int_{\mathbb{S}^{d-1}} \Gamma_{\scS'}(\x,\hv)\bar{U}_{\scS'}(\x|\hv) \hv \, d\hv,
\end{equation} 
and the variance-covariance tensor to
$$\mathbb{D}_{\scS'}(\x)=\dfrac{1}{\displaystyle\int_{\mathbb{S}^{d-1}}\Gamma_{\scS'}(\x,\hv)d\hv}
\displaystyle \int_{\mathbb{S}^{d-1}} \Gamma_{\scS'}(\x,\hv)
\bar{D}_{\scS'}(\x|\hv) \hv\otimes\hv \, d\hv-
{\bf U}_{\scS'}(\x) \otimes {\bf U}_{\scS'}(\x)\,,$$
where $\bar U_{\scS'}$ and $\bar D_{\scS'}$ are respectively given by Eqs.\eqref{barU} and \eqref{barD}.




The fact that $R_{\scS'}^{\scM'}$ and therefore $\Gamma_{\scS'}$ may depend on $\hv$, in general  leads to different speeds in different directions. This means that, unless for very special cases, the 0$^{th}$-order of Eq.\eqref{U0.rp} does not vanish identically and therefore the proper scaling is hyperbolic.
So, even if there is no directional sensing, in presence of a barrier there may be anisotropy due to a non-homogeneous sensing possibility of the cell in the different directions.


\subsection{Volume filling}\label{volumefilling}

In classical volume filling models the speed substituted in the mass balance equation or in the advection-diffusion equation  is a decreasing function of the density, eventually vanishing for densities above a critical value $\rho_{th}$. To handle a similar case in the kinetic framework, we consider $\cS'=\cM'=\rho$ and $\cM_{th}=\rho_{th}$. 
For example, volume filling effects can be described by a $\psi$ with mean 
\begin{equation}\label{vf.old}
\bar{v} (\x|\rho)= \bar v_M \left( 1-\frac{\rho}{\rho_{th}}\right)_+,
\end{equation} 
where $(\cdot)_+=\max(\cdot,0)$ represents the positive part operator and $\bar{v}_M$ is the maximal speed.
For sake of simplicity, we will call $R_{\rho}$ the quantity $R_{\scS'}^{\scM'}$ defined in Eq.\eqref{RMS} and in the following equation we explicitly stress that in presence of volume filling effects the sensing radius $R_{\rho}$ depends on time because the density of cells depends on the evolution of the distribution function $p$ itself. 

\begin{figure}
\begin{center}
\includegraphics[width=0.7\textwidth]{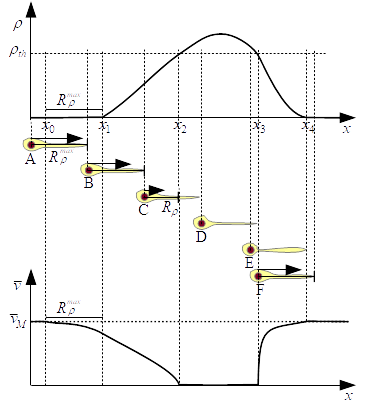}
\end{center}
\caption{Qualitative behaviour of cell speed (bottom) in a landscape with a density that might overcome the volume filling threshold (top). The length of the arrow close to the cells indicates the magnitude of mean speed, while the blockhead arrow the sensing radius $R_\rho$ depending on the presence of the overcrowded area.}
\label{Fig_VF}
\end{figure}

In order to understand the application of the model, let us consider a density distribution like the one in the top of Fig. \ref{Fig_VF} and for sake of simplicity with respect to the general discussion on the maximum sensing radius done at the end of Section 3, let us take $\bar{\varepsilon}=\Delta=0$.
In the discussion we initially take $\gamma_{\rho}(\lambda')=H  (R_{\rho}-\lambda')$, where $H$ is the Heaviside function, meaning that the speed is determined by uniformly averaging the response to the signal $\rho$ in the direction $\hat\vb$ looking ahead up to a distance $R_{\rho}$ defined by \eqref{RMS}. We will finally assume that all cells want to move to the right.
 
Referring to Figure \ref{Fig_VF}, we can qualitatively observe that cell A has all the available space to move and will do it with the maximum allowed speed, up to when it reaches the location $x_0$ because then it is feeling an overcrowded environment ahead. Because of that, it will then slow down stopping upon reaching the point $x_2$. We observe that cell C has a sensing radius limited by the presence of the overcrowded region ahead $x_2$. Cells D and E can not move because they are in the middle of the jammed area ($R_{\rho}=0$). Actually, cell E can sense that there would be available space beyond the point $x_3$, but even is it is close to the border of the jammed area it is still.
On the other hand, cell F can move to the right because it perceives available space ahead. Conversely, if it were polarized to go to the left, then it would not move because it is at the border of an overcrowded region ahead. This clearly points out that the speed of cell F depends on its polarization, an information that in the model is introduced starting from the dependence of $R_\rho$ from $\hv$. From a qualitative point of view, the resulting speed of a cell moving to the right in the landscape given on the top of Fig. \ref{Fig_VF} is given at the bottom of the same figure.

If, instead, $\gamma'_{\scr} (\lambda')=\delta(\lambda'- R_{\scr} )$ cells will only look at a distance $R_{\scr} $ ahead without considering the information within or beyond that distance.  So, in the landscape on the top of Figure \ref{Fig_VF} cell A will move at the maximum speed, cell B at a lower speed (even lower than in the case of a Heaviside function, because it is not averaging over the sensing region) and will stop upon reaching the point $x_2-R_{\scr} ^{max}$. Cells C, D and E will not move, while cell F will start moving. The behaviour of cells B and C could be in principle justified by a will of stopping at a distance from the overcrowded region. As we shall see, because of this effect choosing a Dirac delta as sensing weight will give rise to patterns often characterized by a typical wavelength of the order of  $R_{\scr} ^{max}$.

In this case the transition probability \eqref{TSR0half} writes
\begin{equation}
T[\rho](t,\x,v,\hat{\vb})=\dfrac{1}{\displaystyle \int_{\mathbb{S}^{d-1}}\Gamma_{\scr}(t,\x,\hv)\, d\hv }\int_0^{R_{\rho}(t,\x,\hat\vb)}\gamma_{\rho}(\lambda')\psi(\x,v|\rho(t,\x+\lambda'\hat\vb))\,d\lambda' .
\end{equation}

In particular, we recall that because of the definition (\ref{RMS}), if $\Delta=0$, $R_{\scr} (t,\x,\hat{\vb})$ is such that
\begin{equation}\label{R_rho}
[0,R_{\rho}(t,\x,\hat\vb)]=
\left\lbrace\lambda'\in[0,R_\rho^{max}] \bigg|  \rho(t,\x+\lambda'\hv) \leq \rho_{th} \right\rbrace\, .
\end{equation}
Therefore, in the integration interval $[0,R_\rho(t,\x,\hv)]$
$$\bar{v} (t,\x|\rho(t,\x+\lambda'\hv))= \bar v_M \left( 1-\frac{\rho(t,\x+\lambda'\hv)}{\rho_{th}}\right)\, .$$ 
The mean speed will then be
\begin{equation}\label{bar_U_rho}
\bar U_{\rho} (t,\x|\hv)=\bar{v}_M\left( 1-\dfrac{\bar{\rho}(t,\x,\hv)}{\rho_{th}}\right),
\end{equation}
where
\begin{equation}\label{rhobar}
\bar{\rho}(t,\x,\hv)= \dfrac{\displaystyle\int_0^{R_\rho(t,\x,\hv)}
\gamma_\rho(\lambda')\rho(t,\x+\lambda'\hv) \,d\lambda'}{\Gamma_{\scr}(t,\x,\hv)}\,,
\end{equation}
measures the average density in the direction $\hv$ until the threshold value is reached.


Looking at the macroscopic speed, we have that 
\begin{equation}\label{U.macro.vf}
\Ub_\rho(t,\x)=
\dfrac{\bar{v}_M}{\displaystyle\int_{\mathbb{S}^{d-1}}\Gamma_{\scr} (t,\x,\hv)\,d\hv}
\left(\displaystyle \int_{\mathbb{S}^{d-1}} \Gamma_{\scr} (t,\x,\hv)\hv \, d\hv-
\frac{1}{\rho_{th}}
\displaystyle \int_{\mathbb{S}^{d-1}}  \Gamma_{\scr} (t,\x,\hv)\bar{\rho}(t,\x,\hv)\, \hv \, d\hv \right),
\end{equation}
and it vanishes in points where the density is above the threshold value.

It is instructive to examine the limit case of small $R_{\rho}^{max}$, which implies that $\lambda'\le R_{\rho}^{max}$ is small as well. Like in Section 4 we can perform a Taylor expansion of $\rho(\x+\lambda'\hv)$ in a small neighbourhood of $\x$ with size $R_{\rho}$
\begin{equation}\label{rho_dev}
\rho(t,\x+\lambda'\hv)\approx\rho(t,\x)+\lambda'\nabla \rho(t,\x)\cdot \hv \, .
\end{equation}
Therefore, up to re-scaling,
$$
\bar{U}_{\rho}(\tau,\boldsymbol{\xi}|\hat{\vb})=
\bar{v}_M\left(1-\dfrac{\rho(\tau,\boldsymbol{\xi})}{\rho_{th}} -\epsilon\dfrac{\Lambda'}{\rho_{th}}
\hv\cdot\nabla \rho(\tau,\boldsymbol{\xi})\right)\,,
$$
where
$$
\Lambda'(\tau,\boldsymbol{\xi},\hv)=\dfrac{1}{\Gamma_{\scr} (\tau,\boldsymbol{\xi},\hv)}
\displaystyle \int_{0}^{R_{\rho}(\tau,\boldsymbol{\xi},\hv)} \gamma_{\rho}(\lambda')\lambda' \, d\lambda'
$$
represents a mean sensing distance.
Hence
\begin{equation}\label{bar_U_0_1}
\bar{U}_{\rho}^0=\bar{v}_M\left(1-\dfrac{\rho}{\rho_{th}} \right), \qquad \bar{U}_{\rho}^1= \dfrac{\Lambda'}{\rho_{th}}
\hv\cdot\nabla \rho.
\end{equation}
Therefore the macroscopic velocities of $0^{th}$- and $1^{st}$-order away from overcrowded areas are, respectively,
\begin{equation}\label{U_0_macro_rho}
\Ub_\rho^0=\bar{v}_M
\dfrac{\displaystyle\int_{\mathbb{S}^{d-1}}\Gamma_{\scr}(\tau,\boldsymbol{\xi},\hv)\, \hv\,d\hv}
{\displaystyle\int_{\mathbb{S}^{d-1}}\Gamma_{\scr}(\tau,\boldsymbol{\xi},\hv)\, d\hv}
\left(1-\dfrac{\rho}{\rho_{th}} \right)\,,
\end{equation}
and
\begin{equation}\label{even.vf.1}
\Ub_\rho^1=-\,\dfrac{\bar{v}_M}{\rho_{th}}\mathbb{T}\nabla\rho\,, \qquad{\rm with}\quad
\mathbb{T}(\tau,\boldsymbol{\xi})=
\dfrac{1}{\displaystyle\int_{\mathbb{S}^{d-1}}\Gamma_{\scr}(\tau,\boldsymbol{\xi},\hv)\, d\hv}
\int_{\mathbb{S}^{d-1}}\displaystyle \Gamma_{\scr}(\tau,\boldsymbol{\xi},\hv)\Lambda'(\tau,\boldsymbol{\xi},\hv)
\hv\otimes\hv\,d\hv\,,
\end{equation}
which is anisotropic if $\Gamma_{\scr}$ is not isotropic.

\begin{figure}[!htbp]
\begin{center}
\subfigure[(a)]{\includegraphics[scale=0.3]{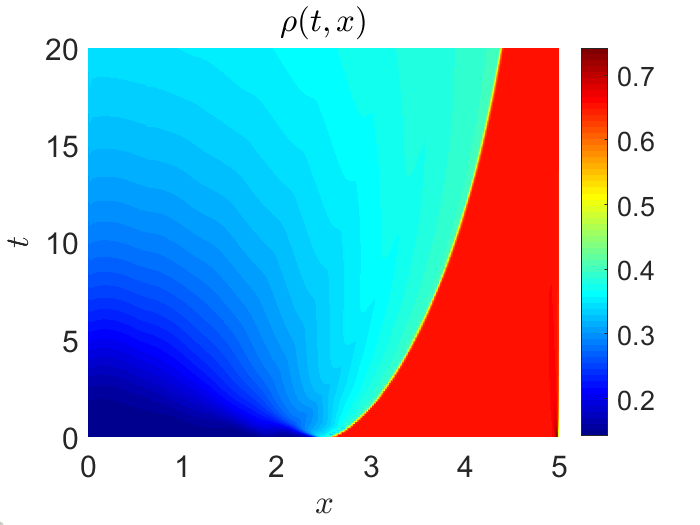}}
\subfigure[(b)]{\includegraphics[scale=0.3]{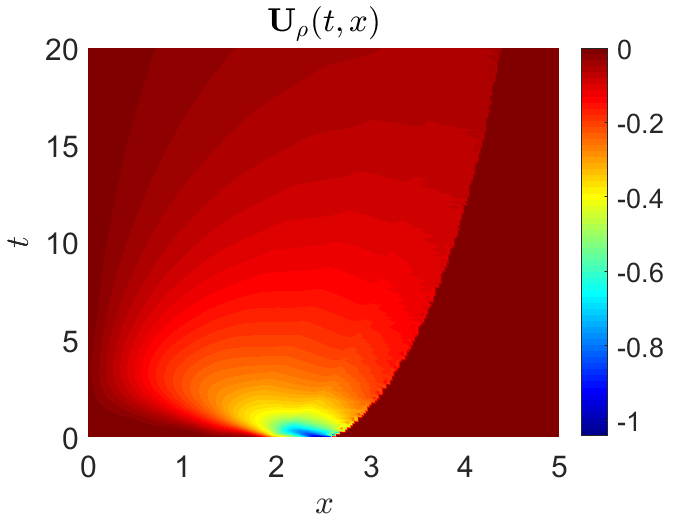}}
\\
\subfigure[(c)]{\includegraphics[scale=0.3]{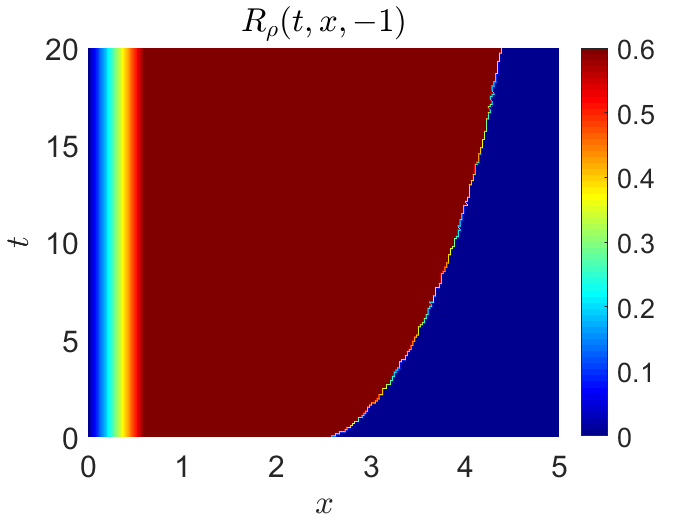}}
\subfigure[(d)]{\includegraphics[scale=0.3]{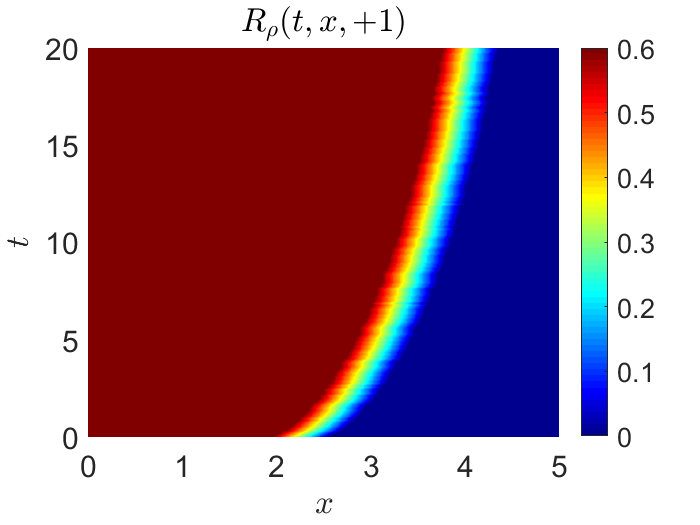}}
\\
\subfigure[(e)]{\includegraphics[scale=0.3]{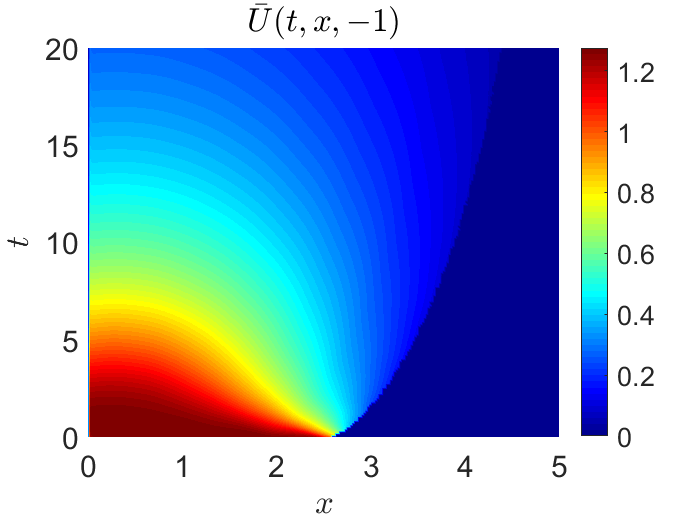}}
\subfigure[(f)]{\includegraphics[scale=0.3]{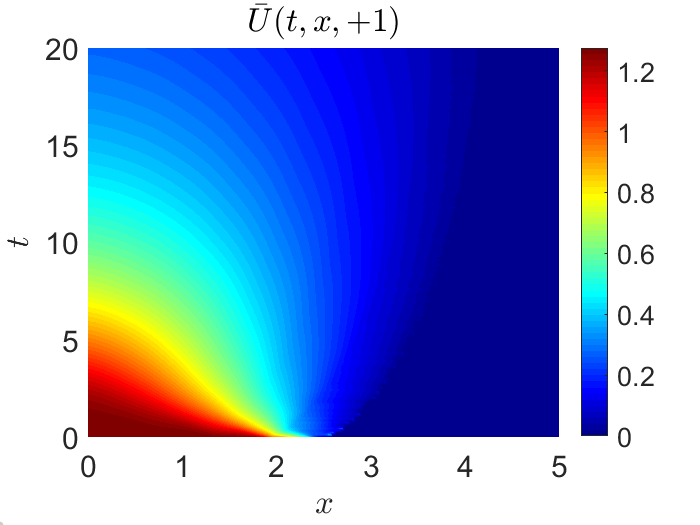}}
\\
\includegraphics[scale=0.3]{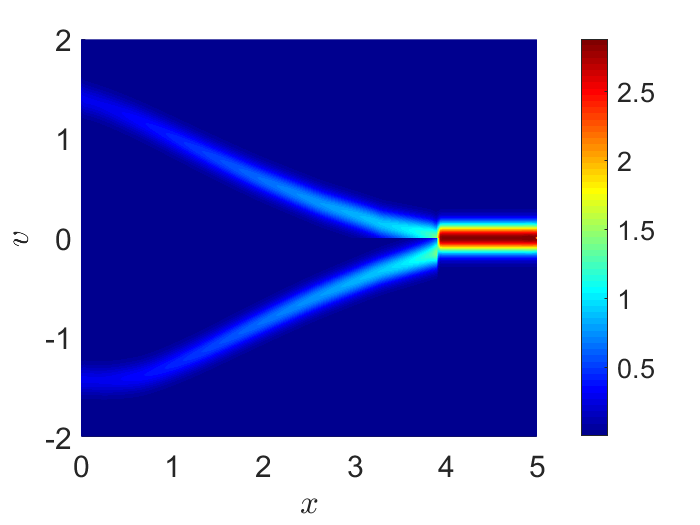}
\includegraphics[scale=0.3]{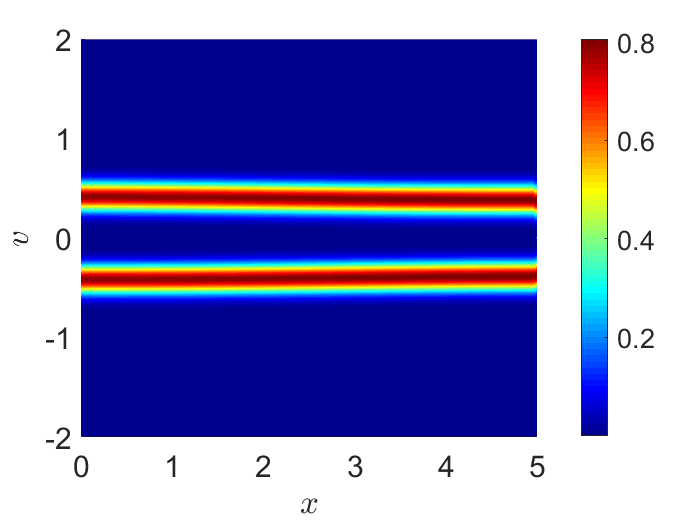}
\end{center}
\caption{Evolution from the initial condition $\rho_0(x)=\frac{0.8}{\pi}\tanh(30(x-2.5))+0.4$, so that $\rho(x)<\rho_{th}=0.4$ if and only if $x<2.5$. The sensing kernel is a Heaviside function. The other parameters are $R_{\scr}^{max}=0.6$ and $\mu=200$. 
In the top row the spatio-temporal evolution of the macroscopic variables $\rho$ and $U_{\scr}$ are plotted. 
In the second row that of the sensing radii in both directions. 
In the third row the mean speed in the two directions and the macroscopic velocity. 
Finally, in the bottom row, the distribution densities $p$ at time $5$ and $10$. Positive and negative $v$'s stand for the speed distributions of cells oriented along $x$ and $-x$.
}
\label{fig.vf.3}
\end{figure}

In this limit, we can perform a parabolic scaling only where $R_{\scr}$ is even or it does not depend on $\hv$, so that the numerator in \eqref{U_0_macro_rho} vanish.
If so, then the parabolic limit reads
$$
\dfrac{\partial \rho}{\partial \tau}-\nabla \cdot \left(\rho \mathbb{T}\nabla\rho \right)=\nabla \cdot \left[ \dfrac{1}{\mu}\nabla \cdot \left(\mathbb{D}_{\rho}^0\rho \right)\right]\,, $$
with
$$
\mathbb{D}_{\rho}^0=\displaystyle \int_{\mathbb{S}^{d-1}} \bar{D}^0_{\rho}(\boldsymbol{\xi}|\hv)\hv\otimes \hv \, d\hv.
$$
Of course, this is barely the case in presence of physical limits of migration (see for example Fig. \ref{fig.vf.3}), and we should perform a hyperbolic limit leading to Eq. \eqref{macro.drift}. 

We can also discuss the proper choice of scaling by considering a nondimensionalization and show that because of the dependence of the sensing radius on the direction, a diffusive time scale can be hardly chosen uniformly in $\Omega$. Therefore, a hydrodynamic limit will be the appropriate one.
Let us now introduce $l_{\scr}=\max \dfrac{\rho}{|\nabla \rho|}$ as the characteristic length of variation of $\rho$, and the parameter
\begin{equation}\label{eta}
\eta=\dfrac{\bar{R}_{\scr}}{l_{\scr}},
\end{equation}
where $\bar{R}_{\scr}$ is a reference sensing radius.
We have seen that if  $\eta \ll 1$, then we may write Eqs. \eqref{rho_dev}, \eqref{bar_U_0_1}, \eqref{U_0_macro_rho}, and Eq. \eqref{even.vf.1}. This is not possible if $\eta \gg 1$.    
We shall rescale the variables as
$$
\boldsymbol{\xi}=\dfrac{\x}{l_{\rho}}, \quad \tilde{v}=\dfrac{v}{U_{ref}}, \quad\tau=\dfrac{t}{\sigma_t}, 
$$
The time scale $\sigma_t$ can be chosen as a diffusion time $T_{Diff}$ scale or a drift time scale $T_{Drift}$. In general we may write \citep{Othmer_Hillen.00}
$$
T_{Diff}=\dfrac{l_{\rho}^2}{D},\qquad D=\dfrac{U_{ref}^2}{\bar{\mu}}, \qquad T_{Drift}=\dfrac{l_{\rho}}{U_{ref}}
$$
The regime is diffusive - and we will choose $\sigma_t=T_{Diff}$ - if the frequency $\bar{\mu}$ is very large with respect to the reference time scale $\sigma_t$, $\ie$ if we can find a small parameter $\epsilon$ such that 
\[
\bar{\mu}=\dfrac{\epsilon^{-2}}{\sigma_t}.
\]
The latter is equivalent to
\begin{equation}\label{l_rho}
l_{\rho}=\mathcal{O}\left(\dfrac{U_{ref}}{\epsilon}\right)
\end{equation}
which implies the hierarchy
\begin{equation}
T_{Drift} \ll T_{Diff}. 
\end{equation}\label{hierarhy}
In the present case this is equivalent to 
\[
\eta=\dfrac{\bar{R}_{\scr}}{l_{\scr}} <1.
\]
On the other hand, the macroscopic regime is hyperbolic, and we choose a faster time scale if
\[
\bar{\mu}=\dfrac{\epsilon^{-1}}{\sigma}.
\]
In this case  the appropriate choice will be 
\[
\sigma_t=T_{Drift}=\dfrac{l_{\rho}}{U_{ref}}=\dfrac{\eta}{\bar{\mu}}
\]
as the hierarchy \eqref{hierarhy} does not hold anymore. Hence, the following relation holds
\[
\eta \sim \dfrac{1}{\epsilon} \gg 1.
\]
We observe that the choice $U_{ref}=\bar{U}_{\scr}$ would make the nondimensionalization depend on the direction. Therefore,
we shall choose as in \citep{Loy_Preziosi}
\[
U_{ref}=\bar{R}_{\scr}\bar{\mu}
\]
being $\bar \mu$ a reference frequency.  The same holds true for $\bar{R}_{\scr}$ that can not be chosen equal to $R_{\scr}$ as the latter depends on the direction. On the other hand, in the present case, we can not even consider $\bar{R}_{\scr}=R_{\scr}^{max}$ in the choice of $U_{ref}$ like in \citep{Loy_Preziosi}, because it varies considerably in time and space and it may also be different at a fixed point in space as it depends on the direction. In conclusion, the relation \eqref{l_rho} cannot be considered everywhere in $\Omega$ and, therefore, the choice of a diffusive time scale is not the proper one. 

Referring to Section \ref{Numerics} for details concerning the numerical integration of the kinetic model, we will now
present some simulations focusing on how the model deals with the volume filling effect. In particular, aiming at checking the effect of limiting the sensing radius $R_{\rho}$ because of overcrowding, we will perform simulation with $R_\rho=R_\rho^{max}$ regardless of the presence of thresholds and limiting it because of the threshold. For sake of simplicity, we will refer to the former case as the {\it unlimited $R$ model}.
Of course, as expected, the effect becomes visible when cell density gets closer to the threshold value $\rho_{th}$. 
A second aim is to put in evidence the difference between using a Dirac delta and a Heaviside function as weight function to evaluate the cell density.

\begin{figure}[!htbp]
\begin{center}
\subfigure[]{\includegraphics[width=0.3\textwidth]{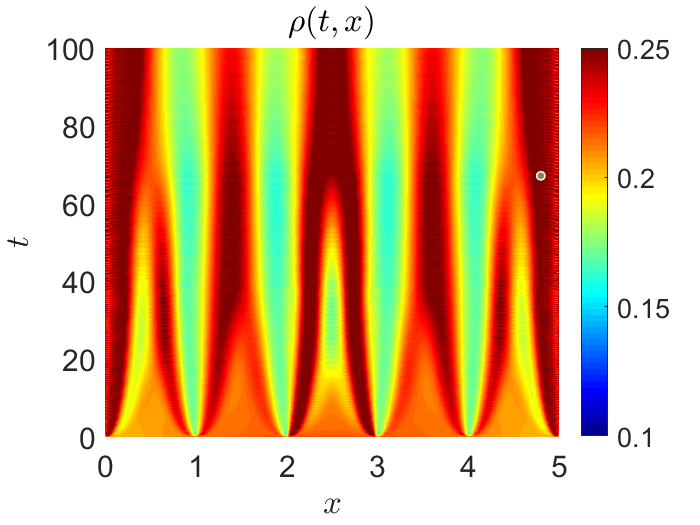}}
\subfigure[$R_\rho=R_\rho^{max}$]{\includegraphics[width=0.3\textwidth]{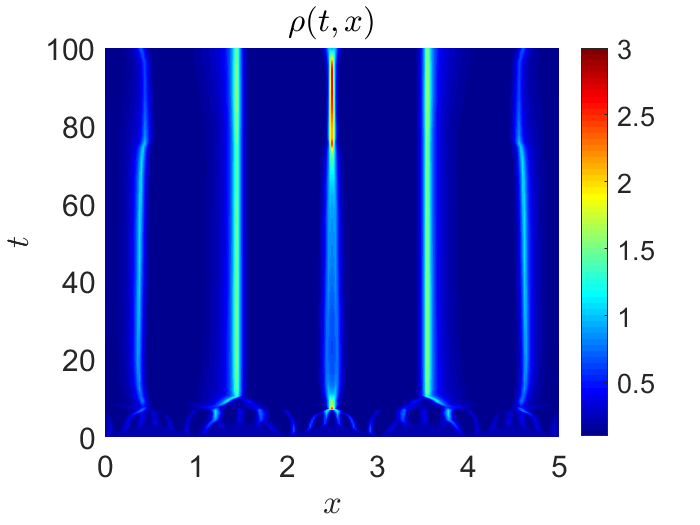}}
\subfigure[$R_\rho=R_\rho^{max}$]{\includegraphics[width=0.3\textwidth]{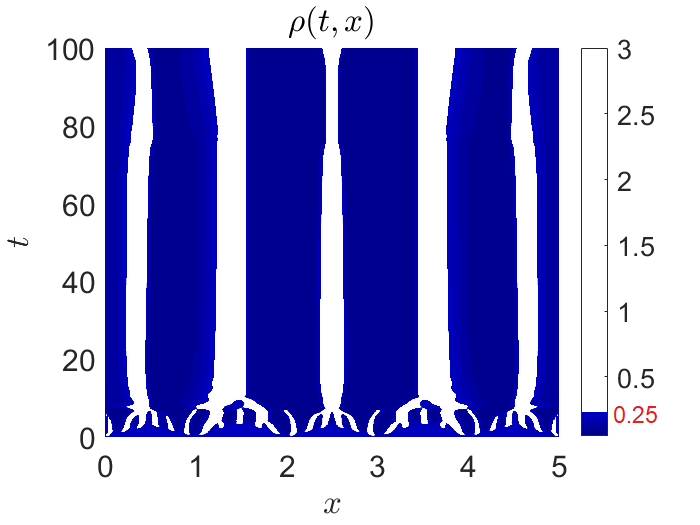}}
\end{center}
\caption{
Pattern formation when $\gamma_{\scr}$ is a Dirac delta starting from a density distribution $\rho_0(x)=0.2+0.02\sin\frac{\pi}{5}  x$ below the threshold value $\rho_{th}=0.25$  and $\mu=10$. In (b) the sensing radius is always $R_{\scr} ^{max}=2$ while in (a) it is limited by $\rho_{th}$. Figure (c) is the same as (b), but the density over the threshold value $\rho_{th}=0.25$ is highlighted in white. }
\label{fig.vf.1}
\end{figure}

In Fig. \ref{fig.vf.1} the initial macroscopic density is a small perturbation of the constant distribution, so that it is always below the threshold value $\rho_{th}$. This implies that  initially $R_{\scr}=R_{\scr} ^{max}$ everywhere. However, the non-homogeneous distribution of speed leads to the formation of overcrowded areas and therefore to the limitation of 
$R_{\rho}^{max}$ in Fig. \ref{fig.vf.1}(a). We observe that the model hampers the cell density from going above $\rho_{th}$. On the contrary, the unlimited $R$ model does not succeed in imposing such a limit (Fig. \ref{fig.vf.1}(b)) and cells go over the threshold value (Fig. \ref{fig.vf.1}(c)).

\begin{figure}[!htbp]
\begin{center}
\subfigure[$\gamma_{\scr} =H$]{\includegraphics[scale=0.26]{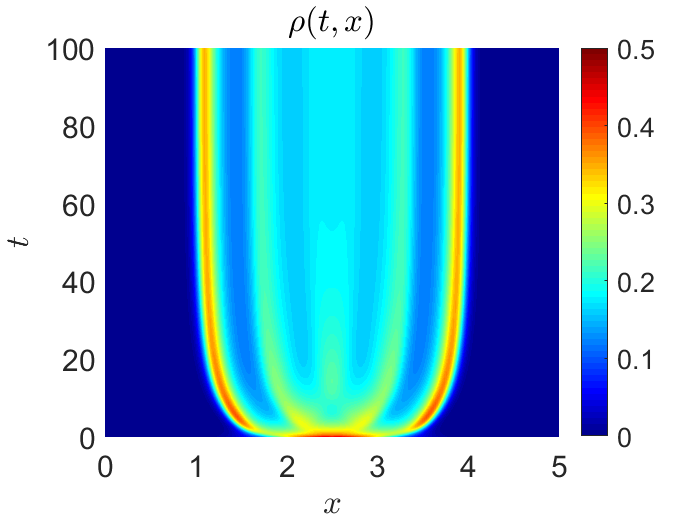}}
\subfigure[$\gamma_{\scr} =\delta$]{\includegraphics[scale=0.26]{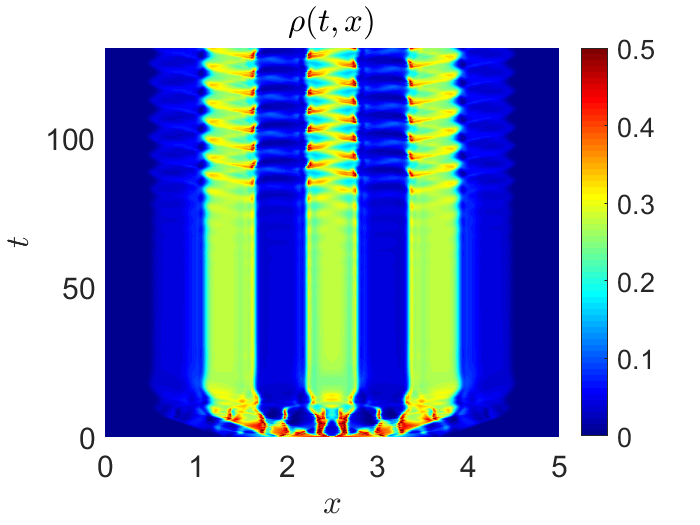}}
\subfigure[$\gamma_{\scr} =\delta$, $R_\rho=R_\rho^{max}$]{\includegraphics[scale=0.26]{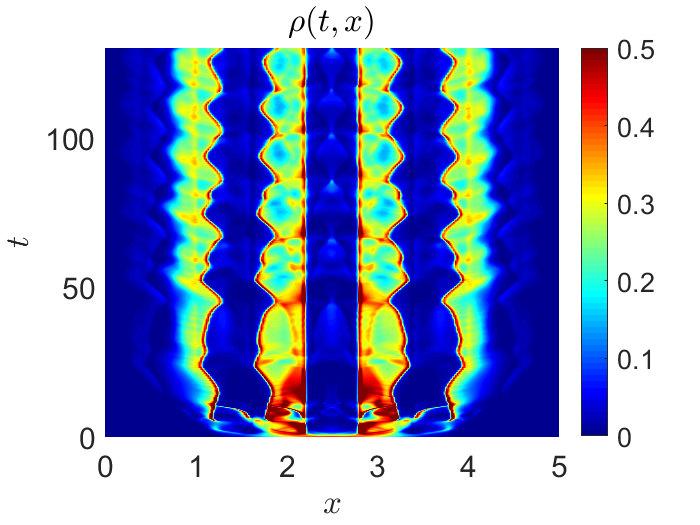}}
\end{center}
\caption{
Pattern formation starting from a density distribution $\rho_0(x)=0.5\exp\left[-\frac{(x-2.5)^2}{0.9}\right]$ 
that just reaches the threshold value $\rho_{th}=0.5$. In addition  $\mu=200$.
In (a) the sensing kernel is a Heaviside function, while in (b) it is a Dirac delta. In addition, in (c) the sensing radius is not limited by 
$\rho_{th}$ and is always $R_{\scr}^{max}=0.6$.}
\label{fig.vf.2}
\end{figure}

In Fig. \ref{fig.vf.2}(a), the sensing function is a Heaviside function. Averaging the cell density over the interval 
$[x,x+R_{\scr}] $, it leads to smoother solutions compared to using a delta function that implies using only the information in $x+R_{\scr}$ in order to determine the new speed (see Fig. \ref{fig.vf.2}(b),(c)). The initial condition barely touches the threshold value $\rho_{th}=0.5$. So, the cell in the center are slower than those away from the center, forming a wave of crowded cells. A second peak forms at a distance $R_{\scr}^{max}$ because cells there sense the advancing front ahead and slow down. This leads to the formation of a pattern of characteristic size comparable with 
$R_{\scr}^{max}$. The fact that cell density in Fig. \ref{fig.vf.2}(a) never reaches the threshold value implies that $R_{\scr}$ is always equal to the the maximum possible value $R_{\scr}^{max}$.

On the other hand, when the sensing function is a $\delta$, the pattern is stronger and is characterized by  maxima that reach $\rho_{th}=0.5$. We also observe a zig-zag behaviour which is a characteristic of other alignment-repulsion-attraction models like in the works by \cite{Eftimie} and \cite{Eftimie2}. This is due to the following dynamics: referring to Fig. \ref{fig.vf.2}(b) at $t\approx 80$, in a certain point cells start clustering, reaching their maximal local density $\rho_{th}$. So, cell sensing this excessive crowding prefer to move in the opposite direction of motion clustering in turn to  values close to $\rho_{th}$. As in the previous figure using an unlimited $R$ model leads to higher peaks beyond $\rho_{th}$ and sharper fronts.

In Fig. \ref{fig.vf.3} the initial condition is not symmetric with respect to the midpoint of $\Omega$, it is smaller than $\rho_{th}$ for $x<2.5$ and larger for $x>2.5$ (see the bottom of Fig. \ref{fig.vf.3}(a)). In this case we observe that $R_{\scr}$ is initially zero in the overcrowded region $x>2.5$ both for the cells polarized to the right and those polarized to the left (see the bottom of Fig. \ref{fig.vf.3}(c),(d)) and cells stay still (see Fig. \ref{fig.vf.3} (e), (f)). For $x<2.5$, $R_{\scr}$ is close to $R_{\scr}^{max}=0.6$ but for a region close to the  overcrowded region for the cells polarized to the right (Fig. \ref{fig.vf.3}(d)) and for the region close to the left boundary for the cells polarized toward it (Fig. \ref{fig.vf.3}(c)). So, cells close to the interface of the overcrowded region start to move to the left (note the negative mean speed in Fig. \ref{fig.vf.3}(b)) and the region to the right gradually empties up to reach a homogeneous configuration. We also observe that the fact that  
$\Ub_{\scr}$ does not vanish implies that \eqref{even} is not satisfied and a diffusive limit can not be performed.

\begin{figure}[!htbp]
\begin{center}
\subfigure[$\gamma_{\scr} =H$]{\includegraphics[scale=0.36]{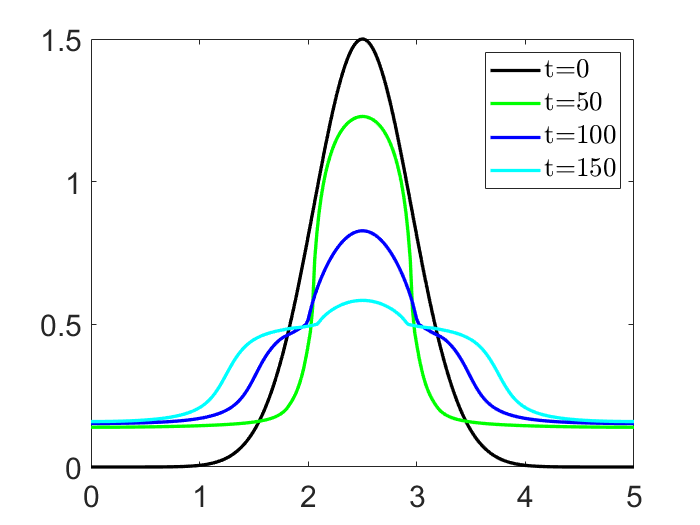}}\\
\subfigure[$\gamma_{\scr} =H$]{\includegraphics[scale=0.36]{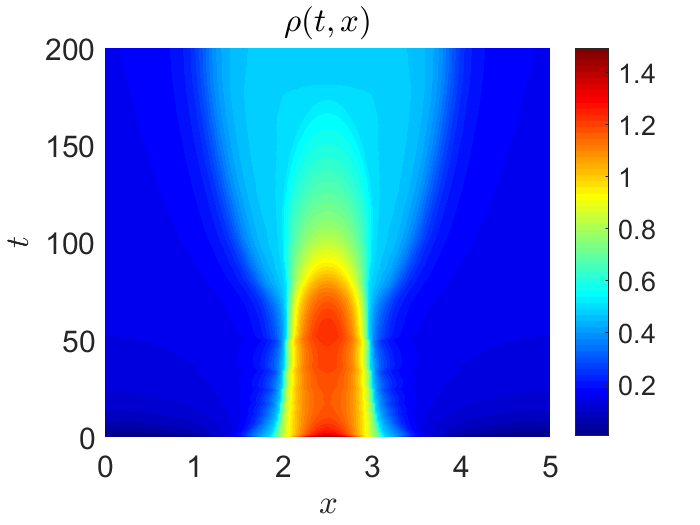}}
\subfigure[$\gamma_{\scr} =\delta$]{\includegraphics[scale=0.36]{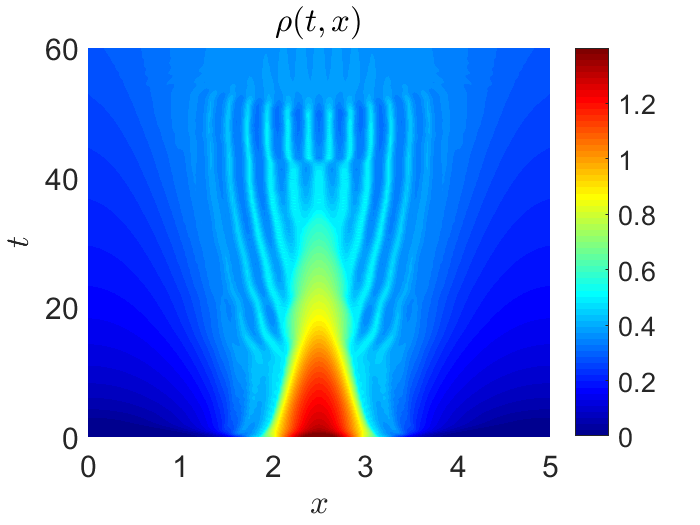}}\\
\subfigure[$\gamma_{\scr} =H$, $R_{\scr}=R_{\scr}^{max}$]{\includegraphics[scale=0.36]{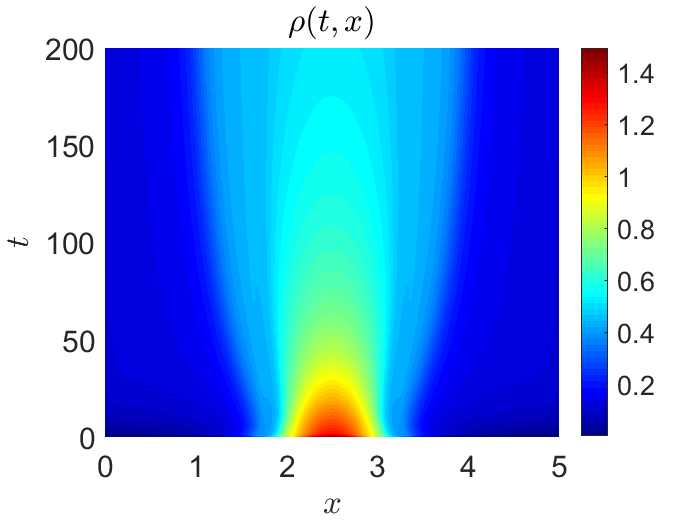}}
\subfigure[$\gamma_{\scr} =\delta$, $R_{\scr}=R_{\scr}^{max}$]{\includegraphics[scale=0.36]{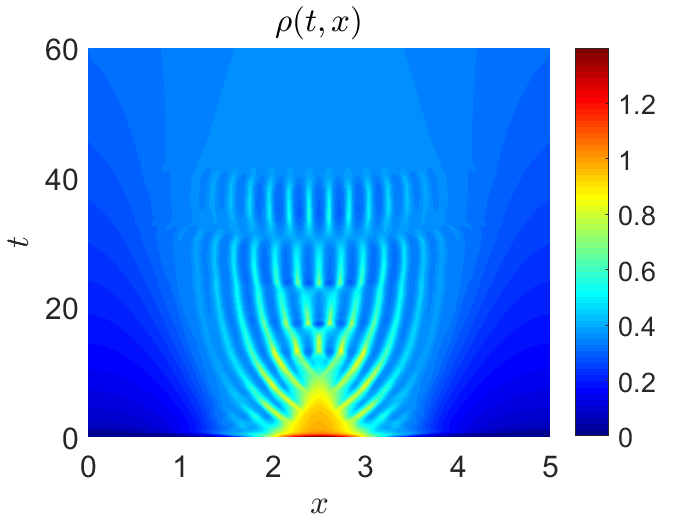}}
\caption{Spatio-temporal evolution from the initial condition $\rho_0(x)=1.5\exp\left[-\frac{(x-2.5)^2}{0.9}\right]$ that has a central region above $ \rho_{th}=0.5$. 
Other parameters are $R_{\scr} ^{max}=0.2$ and $\mu=2$.
In (a,b,d) density for a Heaviside sensing kernel and in (c,e) for a Dirac delta. In particular, in (d,e) 
$R_{\scr}$ is no limited and is always given by $R_{\scr}^{max}$. 
}
\label{fig.vf.4}
\end{center}
\end{figure}

In Fig. \ref{fig.vf.4} the initial distribution is a Gaussian with maximum above $\rho_{th}=0.5$, so  that there is overcrowding in the central region, specifically nearly between $1.8$ and $3.2$. 
In the region where the density is below the threshold, the solution diffuses fast. In fact, cells initially located in $x>3.2$ that want to move to the right will readily do so (see Fig. \ref{fig.vf.4}(a) at time t=100). Then those closer to the border of the above interval move faster than those more in the center, because of the measured nonlocal density, 
leaving back a steeper function (Fig. \ref{fig.vf.4}(a) at time t=100) that then slowly diffuses away.
 If one always takes $R_{\scr}=R_{\scr} ^{max}=0.2$ diffusion is faster (Fig. \ref{fig.vf.4}(d))

The evolutions using a Heaviside sensing kernel (Fig. \ref{fig.vf.4}(a,b,d)) are smoother than those with a Dirac delta (Fig. \ref{fig.vf.4}(c,e)) because the information of the density in the desired direction is averaged, giving rise to smoother velocities, rather than measured at a single point ahead.
In Fig. \ref{fig.vf.4}(c,e) the fact that cells are only considering the cell density at a distance $R_{\scr} ^{max}=0.2$ regardless of whether there are denser regions for lower distances generate a pattern with a characteristic length of the order of $R_{\scr} ^{max}=0.2$ . 
However, since the density is not very high compared to the threshold the patterns fade away.  
The structure of the simulation presented in Fig. \ref{fig.vf.4}(d,e) is similar but faster than  in Fig. \ref{fig.vf.4}(b,c) because more cells measure a density (always at $R_{\scr}^{max}$) allowing motion. The Supplementary movie \textit{VF.mp4}
 shows the time evolution of the density distribution $p$ that starts from a distribution that is uniformly distributed in the velocity space and reaches the spatial uniform distribution.


\subsection{Cell-ECM Interactions}

In order to move in a three-dimensional environment, cells interact with the ECM. This is network of fibres that on one hand are used by cells to adhere and exert traction forces and on the other hand can constitute an obstacle when the characteristic pore size drops below a threshold value. The combination of these interactions leads to a bimodal dependence of cell speed on the stiffness and density of the ECM \citep{Harley, Peyton, Zaman}. In particular, there is a threshold value $M_{th}$ above which cell can not move in the ECM  \citep{Wolf.07, Wolf.11}. In some migration experiments, mainly on artificial scaffolds, it is also clear that when there is little or no possibility of building focal adhesion with the substratum then again cells can not exert active traction forces and are unable to move \citep{Strisce, strisce2}. In terms of ECM we can then infer that there might also be a minimal density $M_0$ of ECM necessary to crawl in it. There is however a difference between the two thresholds because in this last case cells can extend their protrusions beyond the region lacking of adhesion points to possibly reach a farther region where they can adhere and exert traction.

\begin{figure}[!htbp]
\begin{center}
\includegraphics[width=0.8\textwidth]{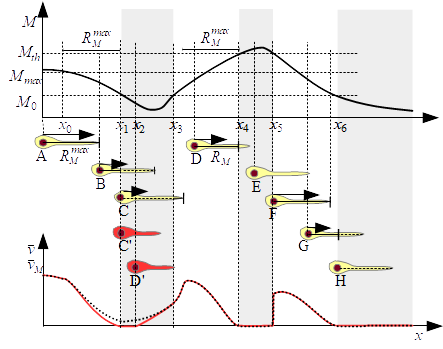}
\end{center}
\caption{
Qualitative behaviour of cell speed (bottom) in a landscape with an ECM density shown above. The grey regions denote problematic regions either because of too dense ECM (the central one) or because of lack of ECM (the lateral ones). 
The length of the arrow close to the cells indicates the magnitude of mean speed, while the blockhead arrow indicates the sensing radius $R_M$ depending on the presence of an area with prohibitively small pore sizes.. The dashed lines close to the cell denote intervals that do not contribute to cell speed. In the bottom graph the black dotted line refers to the mean speed to the yellow (lighter) cell that can extend longer protrusions, while the full red line the one of the red (darker) cell, that has vanishing speed in the interval $[x_1,x_2]$. Both cell types have vanishing speed for $x>x_6$.
}
\label{ECM}
\end{figure}

Denoting by $R_{M} $ the quantity $R_{\scS'}^{\scM'}$ defined in Eq.\eqref{RMS} for the ECM case and 
referring to the ECM landscape in Fig. \ref{ECM}, we can identify a cell response similar to overcrowding where there is an ECM density above $M_{th}$. Namely, for cell A the sensing radius $R_{M} =R_{M} ^{max}$, while for cell D it is  $R_{M} <R_M^{max}$ because cell speed is set according to the ECM density up to $x_4$. Cell E can not move because it is in the middle of the dense ECM region, while cell F can move to the right because it will encounter a microstructure allowing cell motion, but not to the left because in that direction the pore size is too small.

However, as already stated, here a new phenomena occurs that is related to the existence of regions with scarse presence of ECM. In Fig. \ref{ECM} it occurs for $x_1<x<x_3$ and $x>x_6$. In these situations a cell has no limitation of the sensing radius $R_M^{max}$, but when it extends its protrusions it is not able to build focal adhesion and exert  traction forces in these interval, so the related contribution to cell speed vanishes. As a consequence, typically the speed  decreases, as for cell B. Now two situations may occur when a cell reaches this problematic area. If it is able to extend its protrusions to find a place to anchor and adhere, like cell C, it can exert traction forces and jump beyond the interval $[x_1,x_3]$. On the other hand, if it is not able to do so, like the red cell C' that is characterized by a smaller $R_M^{max}$, then it is stuck in $x_1$. Finally, a red cell located at a distance $R_M^{max}$ from $x_3$, like cell D', barely touches the border in $x_3$. So, the interval $[x_1,x_2]$ is characterized by a vanishing speed as $[x_4,x_5]$ but for different dynamics, that will be shown in the simulations to follow. 
For a similar reason the region beyond $x_6$ can not be reached by both the yellow cell H and a red cell in the same location.

In order to take into account of all the effects mentioned above, in the simulations to follow we will use the following specific form of
$$\bar v=\bar v_0+\dfrac{4(\bar v_M-\bar v_0)}{(M_{th}-M_0)^2} [(M-M_0)(M_{th}-M)]_+,$$
representing (if $\bar v_0=0$) the positive part of a parabola with zeros in $M_0$ and $M_{th}$ with maximum speed $\bar v_M$ achieved for $M=\frac{1}{2}(M_{th}+M_0)$ denoted by $M_{max}$ in the following, for sake of simplicity.   
The discussion of the macroscopic limit closely follows what presented for the volume filling case. For this reason it will not be repeated here.

\begin{figure}[!htbp]
\begin{center}
\includegraphics[scale=0.5]{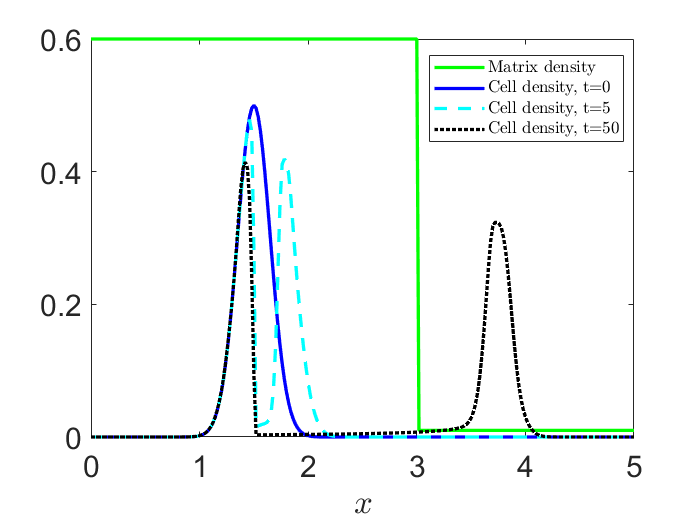}

\end{center}
\caption{Behaviour of a cell distribution initially within a dense ECM area when $R_M$ is not limited as defined by \eqref{R_M}, but is always equal to $R_M^{max}=1.5$. The sensing kernel is a Dirac delta. Other parameters are 
$M_{0}=0$, $M_{th}=0.4$, $\bar v_M=0.7$. }
 \label{ecm.1}
\end{figure}

With the aim of showing the importance of considering a sensing radius defined as in \eqref{R_M}, in Fig. \ref{ecm.1} we present a simulation where the value $M_{th}=0.4$ is smaller than that of the matrix density where cells are initially  located. However, half of the cell population is at a distance that is smaller than $R_{M} ^{max}$ from the border of the dense area (that is in $x=3$). So, if $R_M$ is not limited as defined in \eqref{R_M}, they sense beyond the physical barrier and manage to come out of the dense zone of ECM because they have a positive speed in the direction going out of the ECM. Conversely, if \eqref{R_M} is used cells are blocked in the dense area on the left. 

\begin{figure}[!htbp]
\begin{center}
\subfigure[$R_{M} ^{max}=0.2$]{\includegraphics[scale=0.36]{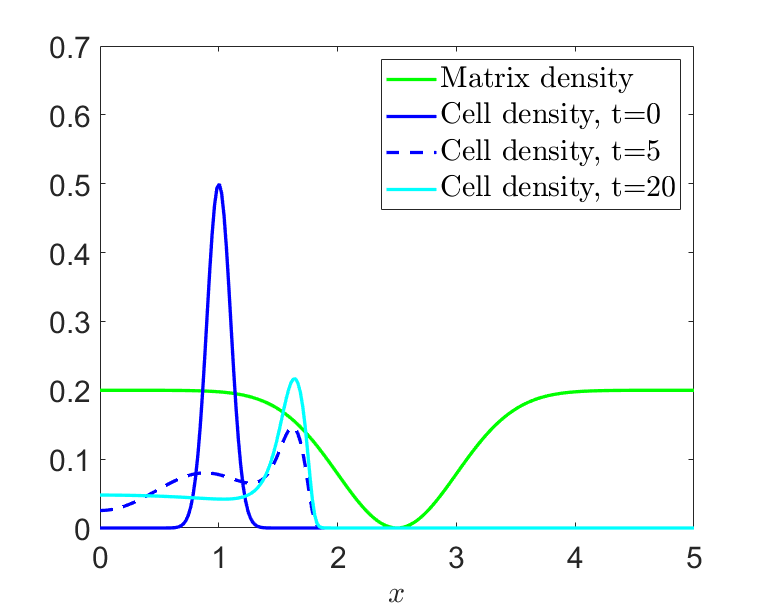}}
\subfigure[$R_{M} ^{max}=2$]{\includegraphics[scale=0.36]{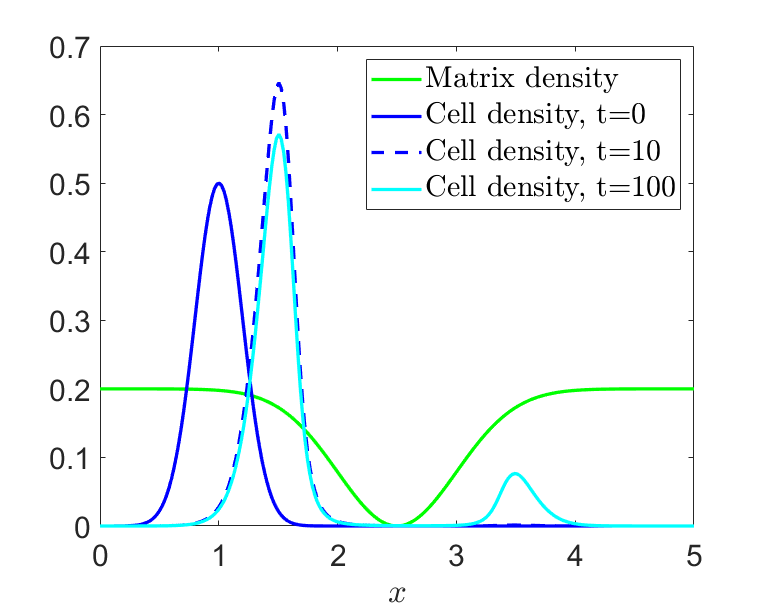}}\\
\subfigure[$R_{M} ^{max}=0.2$]{\includegraphics[scale=0.4]{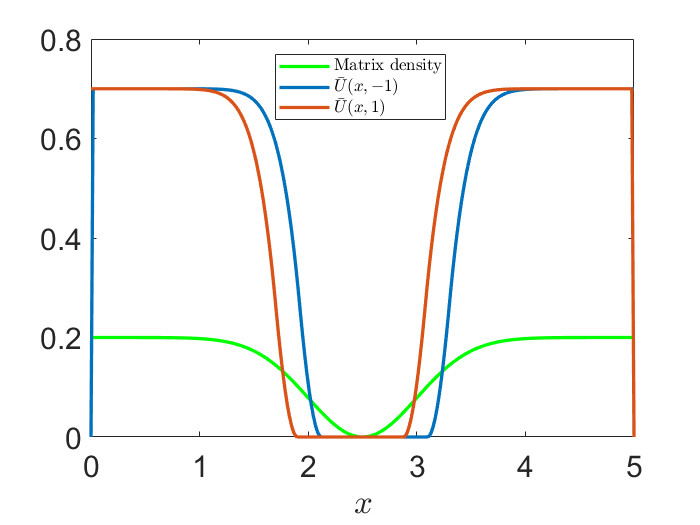}}
\subfigure[$R_{M} ^{max}=2$]{\includegraphics[scale=0.4]{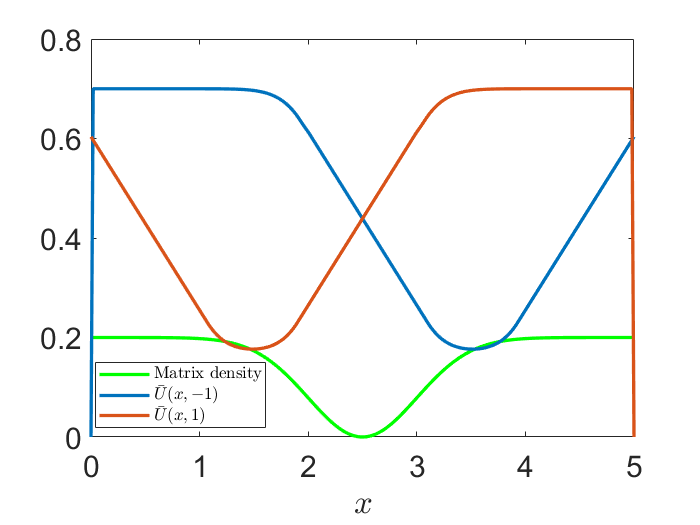}}\\
\subfigure[$R_{M} ^{max}=0.2, p$]{\includegraphics[scale=0.5]{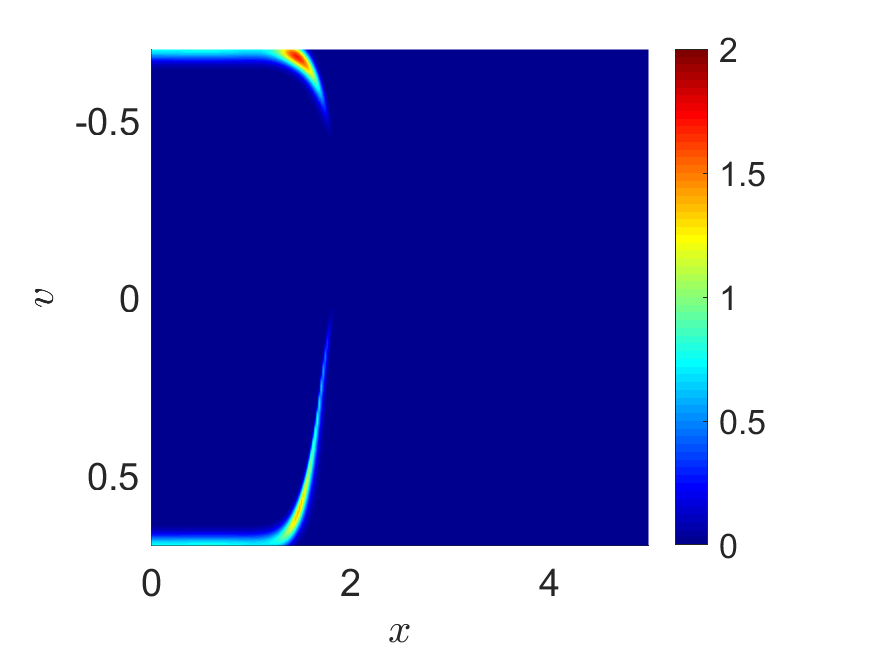}}
\subfigure[$R_{M} ^{max}=0.2, \dfrac{p}{\rho}$]{\includegraphics[scale=0.5]{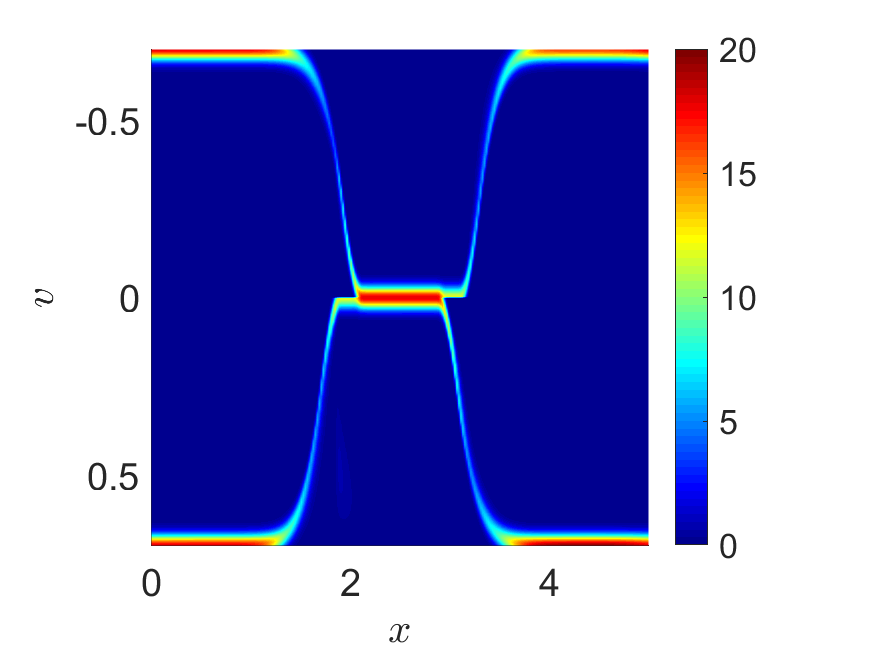}}
\end{center}
\caption{Spatio-temporal evolution of the cell density (a,b) in presence of a region with extremely low density of ECM. The sensing kernel is always a Heaviside function and in (c,d) the mean speed is given with $M_0=0.1$, $M_{th}=0.3$, $\bar v_M=0.7$, $\mu=20$. In (a), (c), (d) and (e) $R_M^{max}=0.2$, while in (b) and (d) $R_M^{max}=2$. In (e)-(f) we present the density distribution $p$ and the probability measure $\dfrac{p}{\rho}$.}
\label{ecm.2}
\end{figure}

In Fig. \ref{ecm.2} we show the effect of the presence of an area with very low density of ECM, a sort of hole in the ECM. In fact, here $M_0=0.1$, but the ECM profile (represented by the green line) presents a central region $[P,Q]$ with density smaller then $0.1$. 
In Fig. \ref{ecm.2}(a), $R_M^{max}=0.2$ and so cells can not go over the point $P$ because, like the red cell in Fig. \ref{ECM}, they can not reach with their protrusions the point $Q$ where the ECM assumes again values larger then $M_0$. 
On the other hand, in Fig. \ref{ecm.2}(b), $R_M^{max}=2$ and, so, cells manage to reach the point $Q$ and go over the hole. In fact, referring to Fig. \ref{ecm.2}(c),(d) for the cells that have short protrusions there is a point where the speed \eqref{barU} vanishes, whilst cells with longer protrusions move with a slower velocity when approaching the ECM depression but always keep a positive speed (Fig. \ref{ecm.2}(d)), so that they manage to overcome the problematic area. In Fig. \ref{ecm.2}(e) the density distribution $p$ shows the distribution of the microscopic velocities in the physical space while in Fig. \ref{ecm.2} (f) $\dfrac{p}{\rho}$ represents the equilibrium probability measure. 

\begin{figure}[!htbp]
\begin{center}
\subfigure[Matrix density]{\includegraphics[scale=0.35]{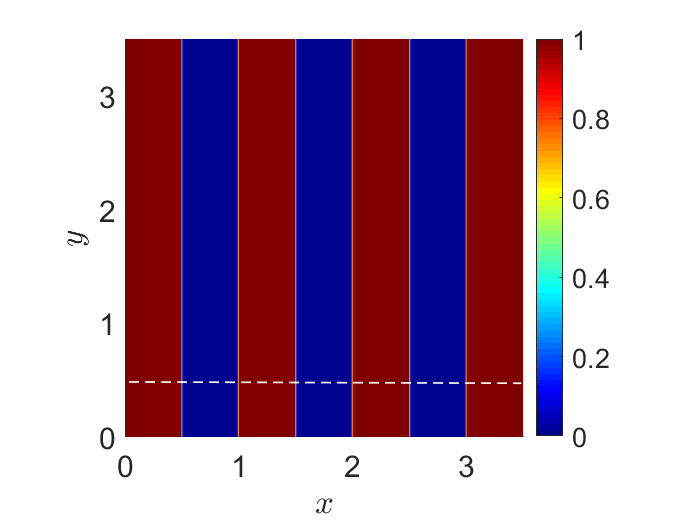}}
\subfigure[$t=1$]{\includegraphics[scale=0.35]{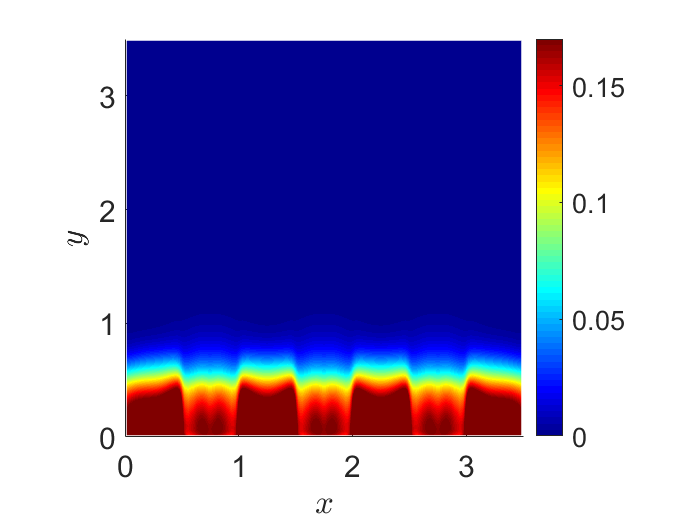}}
\\
\subfigure[$t=10$]{\includegraphics[scale=0.35]{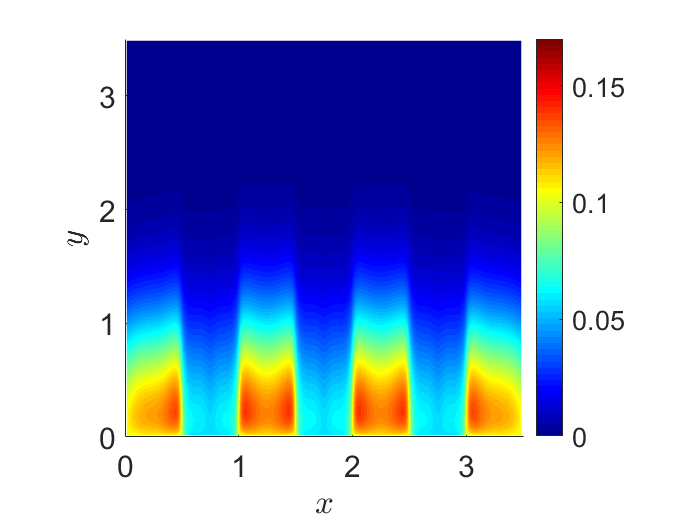}}
\subfigure[$t=35$]{\includegraphics[scale=0.35]{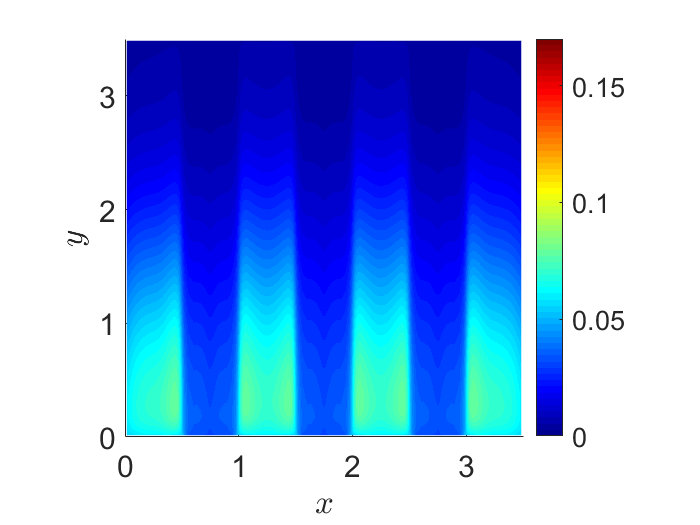}}
\end{center}
\caption{Spatial evolution of cell density in an environment with ECM stripes distributed as in (a). Cells are initially uniformly distributed below the dashed line.  As $M_{0}=0.01$ and $M_{max}=1$, cells prefer to move along the stripes and as $R_{M}^{\max}=0.25$ those located in the middle of the blue stripe initially have some difficulties in reaching them.
The sensing kernel is a Heaviside function and  $\bar v_M=1$. Here the boundary conditions are those defined in \eqref{Maxwell} with $\alpha=0.5$.}
\label{ecm_stripes}
\end{figure}

In Fig. \ref{ecm_stripes} we present a set of two-dimensional simulations mimicking the experimental set-up in which a Petri dish is coated with thick stripes of different ECM components. For instance, in \citep{Strisce} they show the different locomotion on laminin (or E8) and on fibronectin. The stripes made of fibronectin do not encourage locomotion, irrespective of the level of coating concentration used,  whilst  laminin and E8 encourage locomotion. In particular the locomotory response is peaked around a certain range of values of laminin ($20-50 fmol\cdot cm^{-2}$) whilst it decreases for larger and smaller values of coating concentrations. Similar experiments are also performed by \cite{strisce2}. Specifically, as shown in Fig. \ref{ecm_stripes}(a), the density of the laminin coating is $1$ where it is present (red stripes) and zero elsewhere (blue stripes). In this case $M_0=0.01$ and so the cells are not able to adhere in the blue stripes, whilst $M_{max}=1$, and so the cells have a maximal speed on the stripes of laminin. Cells start from an initial uniform distribution in $y<0.5$. 
Cells on the ECM laminin stripes rapidly move along them and avoid going on the stripes lacking of laminin. Cells in the region where there is no laminin cannot move along the blue stripes. However, the sensing radius is sufficiently high to allow all of them to grab some adhesion sites and pull themselves onto the laminin stripes. Initially, (see Fig. \ref{ecm_stripes}(b)) cells closer to the laminin stripes are faster and soon jump onto them to then steer and move along them. This is the reason why there is a minimum of cell density on the blue stripes close to the interface and a faster progression on the red stripes. On the other hand, cells in the center are slower and take longer to move more or less perpendicularly to the stripes. In fact, cells not in the middle of the blue stripe can reach the laminin stripes also if not oriented perpendicularly to them. Eventually, the entire population evolves along the stripes (Fig. \ref{ecm_stripes}(d)). (See Supplementary movie Stripes.mp4). 

\section{Double bias}

We shall now consider the case of cell migration under the action of both a field $\cS$ affecting the choice of the direction of motion (specifically either an external chemoattractant or an internal effect due to cell-cell adhesion) and a field $\cS'$  affecting the speed (specifically related either to volume filling effects or to the presence of ECM). 
In this case both sensing radii may depend on time, position and direction.
We then consider the operator
\begin{equation}\label{nonlocalchemo}
\begin{array}{lr}
\mathcal{J}[p](t,\x,v,\hv)=
\mu (\x) \left(\rho(t,\x)c(\x) \displaystyle \int_{0}^{R_{\scS}(t,\x,\hv)} \gamma_{\scS} (\lambda) b(\cS(\x+\lambda \hv)) \, d\lambda \Psi[\cS']- p(t,\x,v,\hv) \right),
\end{array}
\end{equation}
where $\Psi$ is given by \eqref{Psi}.

\subsection{Adhesion and volume filling}

\begin{figure}[!htbp]
\begin{center}
\subfigure[$\gamma_{\scr,vf}=\delta, R_{\scr, adh}=R_{\scr,vf}$]{\includegraphics[scale=0.45]{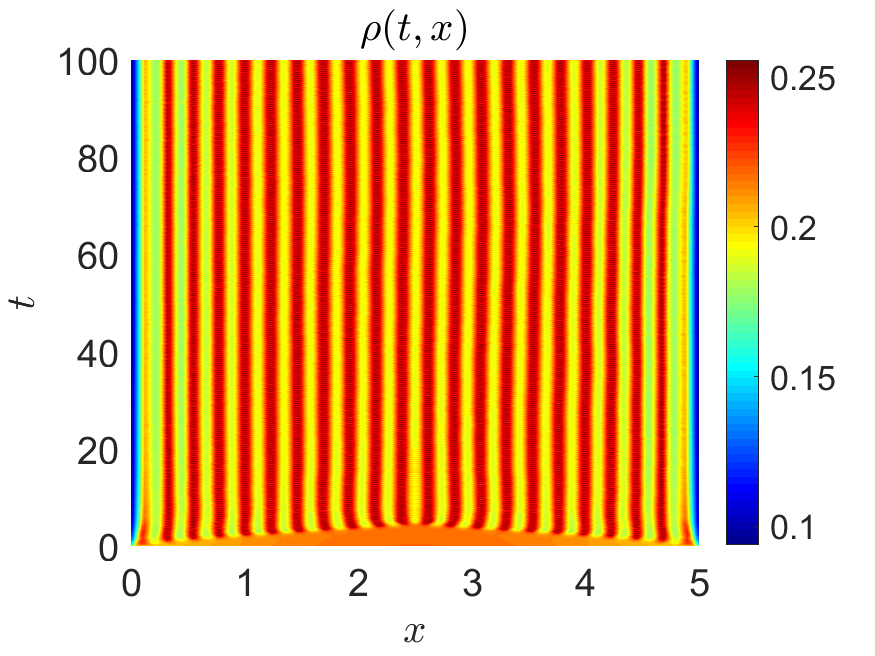}}
\subfigure[$\gamma_{\scr, vf}=\delta,  R_{\scr, adh}=R_{\scr, adh}^{max}$]{\includegraphics[scale=0.35]{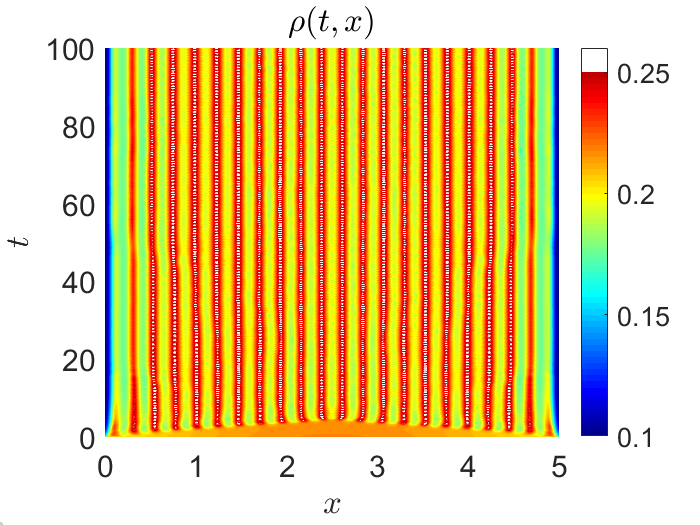}}
\subfigure[$\gamma_{\scr, vf}=H,  R_{\scr, adh}=R_{\scr, vf}$]{\includegraphics[scale=0.35]{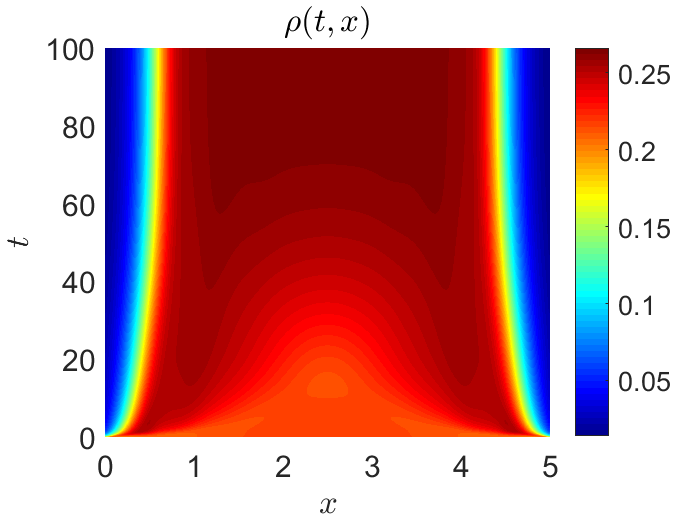}}
\subfigure[$\gamma_{\scr, vf}=H,  R_{\scr, adh}=R_{\scr, adh}^{max}$]{\includegraphics[scale=0.35]{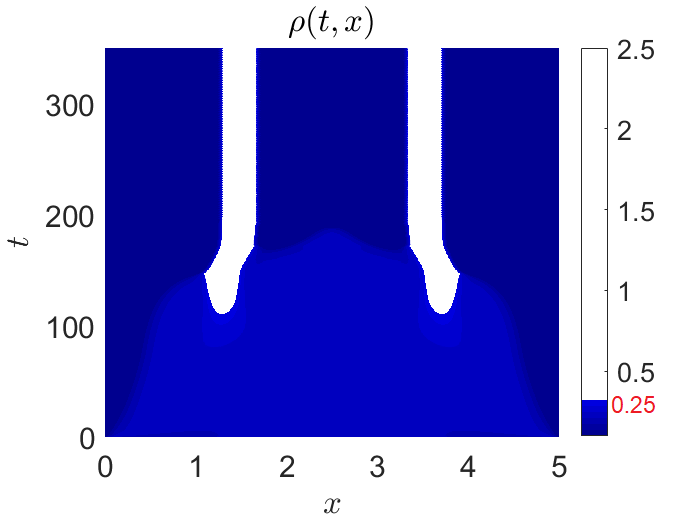}}
\end{center}
\caption{Evolution in presence of cell-cell adhesion and volume filling with $\rho_{th}=0.25$ starting from the initial distribution $\rho_0(x)=0.2+0.02\sin\frac{\pi}{5}x$ with $\mu=2$. The sensing kernel for adhesion is always a Heaviside function, while that for volume filling is a Dirac delta (top row) or a Heaviside function (bottom row). In the left column a limited $R_{\scr}$ is used while in the right column the maximum protrusion length $R_{\scr}^{max}=0.2$. In all figure, densities above the threshold value $\rho_{th}=0.25$ are coloured in white.}
\label{fig.vf.adh.1}
\end{figure}

Let us consider the case in which cell-cell adhesion represents a mechanism of cell polarization biasing the otherwise random motion of cells but taking into account that cells can not come too close because of  volume filling, $\ie$, $\cS=\cS'=\rho$. We shall denote $R_{\scS}=R_{\rho,adh}$ and $R_{\scS'}=R_{\rho,vf}$. In particular, as the two cues are the same it is natural to take $R_{\rho,adh}=R_{\rho,vf}$ and we will denote it by $R_{\scr}$. Figure \ref{fig.vf.adh.1} highlights the differences between when $R_{\scr}$ is defined as in Section \ref{volumefilling} (see (a) and (c)) and when it is equal to the maximum possible extension of protrusions $R_{\scr}^{max}$.
In particular, in Fig. \ref{fig.vf.adh.1}(a,b), we observe that if the sensing function for the volume filling is a Dirac delta the difference is not so remarkable, with the formation of patterns of size close to $R_{\rho}^{max}$. We also highlight the formation of a cell-free zone close to the boundary because of the action of adhesion forces that pull the aggregate together. On the other hand, if the sensing kernel for volume filling is a Heaviside function, the two different choices of sensing radius determine the formation of a plateau as in Fig. \ref{fig.vf.adh.1} (c), or two aggregates as in Fig. \ref{fig.vf.adh.1} (d). Moreover in the latter case, the cell density goes over the threshold $\rho_{th}$. In all the other cases the cell density remains under the threshold value $\rho_{th}$.
%

\subsection{Chemotaxis and volume filling}

In this case we assume that cells are sensitive to a chemoattractant with  $b(\cS)=\cS$ and take into account of volume filling effects with $R_{\scr} (t,\x,\hv)$ given by \eqref{R_rho} and mean speed $\bar{U}_{\scr} (\boldsymbol{\xi}|\hat{\vb})$ by \eqref{bar_U_rho}. 
In general the appropriate macroscopic limit will be a hyperbolic one, which, dropping the dependence from space and time for sake of implicity, reads
\begin{equation}
\dfrac{\partial \rho}{\partial \tau}+\nabla \cdot \left[ \bar v_M \rho\displaystyle\dfrac
{\displaystyle\int_{\mathbb{S}^{d-1}}\Gamma_{\scS}(\hv)\Gamma_{\scr}(\hv)\bar\cS(\hv)\left(1-\frac{\bar\rho(\hv)}{\rho_M}\right)\hv\,d\hv}
{\displaystyle\int_{\mathbb{S}^{d-1}}\Gamma_{\scS}(\hv)\Gamma_{\scr}(\hv)\bar\cS(\hv)\,d\hv}\right]=0
\end{equation}
where $\bar\rho$ is given by \eqref{rhobar} and, analogously,
$$\bar\cS(\hv)=\dfrac{1}{\Gamma_{\scS}(\hv)}\int_0^{R_{\scr} (\hv)}
\gamma_{\scS}  (\lambda) \cS(\boldsymbol{\xi}+\lambda \hv)  \, d\lambda\,$$
is the weighted average of the signal $\cS$.

\begin{figure}[!htbp]
\begin{center}
{\includegraphics[scale=0.5]{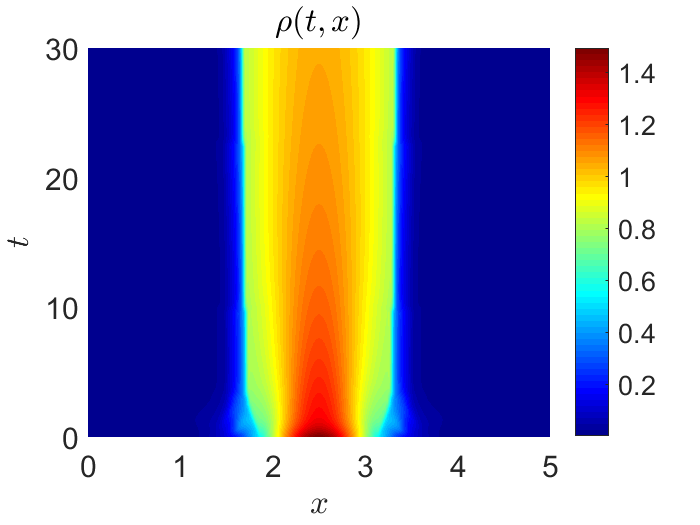}}
\end{center}
\caption{
Spatio-temporal evolution from the initial condition $\rho_0(x)=1.5\exp\left[-\frac{(x-2.5)^2}{0.45}\right]$ as in Fig. \ref{fig.vf.4} but in presence of a chemoattractant  distributed as $\cS(x)=0.3\exp\left[\frac{(x-2.5)^2}{0.02}\right]$. The sensing kernel is a Heaviside function. 
Other parameters are $R_{\scS} ^{max}=R_{\scr}^{max}=0.2$, $\rho_{th}=0.5$, $\mu=2$.
}
\label{fig.vf.ch.1}
\end{figure}

In Fig. \ref{fig.vf.ch.1}, we consider the same volume filling effect as in Fig. \ref{fig.vf.4} under the action of a normally distributed chemoattractant centered in $x=2.5$. 
The sensing radius for the chemoattractant is the same as the one for the volume filling effect defined as in \eqref{R_rho} which is affected by the threshold value $\rho_{th}$. Due to the presence of the chemoattractant, the cell density does  not converge to the constant solution, but it remains above the threshold value as the chemoattractant and the volume filling effect are balanced.

\subsection{Chemotaxis and steric hindrance}

\begin{figure}[!htbp]
\begin{center}
\subfigure[Inital condition]{\includegraphics[scale=0.25]{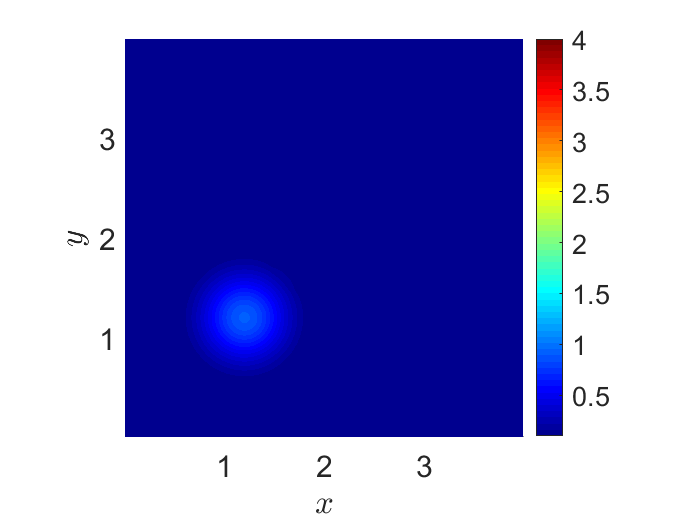}}
\subfigure[Chemoattractant]{\includegraphics[scale=0.25]{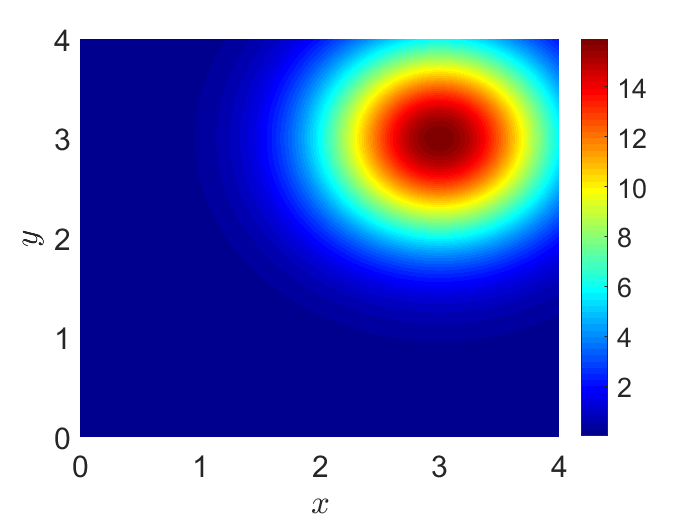}}
\subfigure[Matrix density]{\includegraphics[scale=0.25]{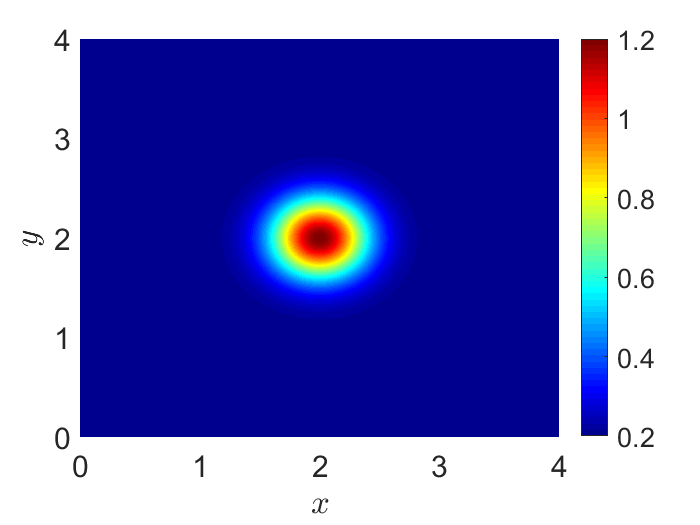}}
\\
\subfigure[$t=5$]{\includegraphics[scale=0.25]{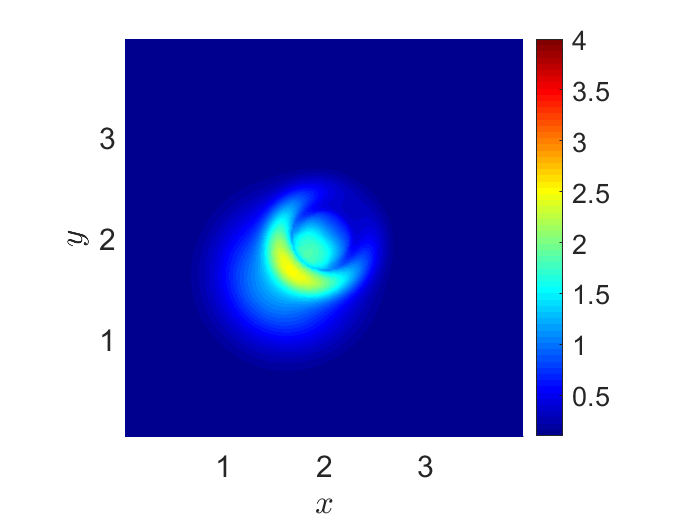}}
\subfigure[$t=10$]{\includegraphics[scale=0.25]{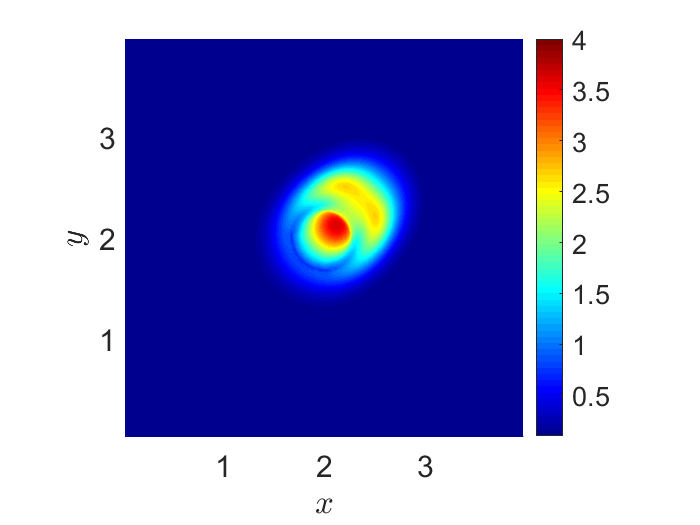}}
\subfigure[$t=50$]{\includegraphics[scale=0.25]{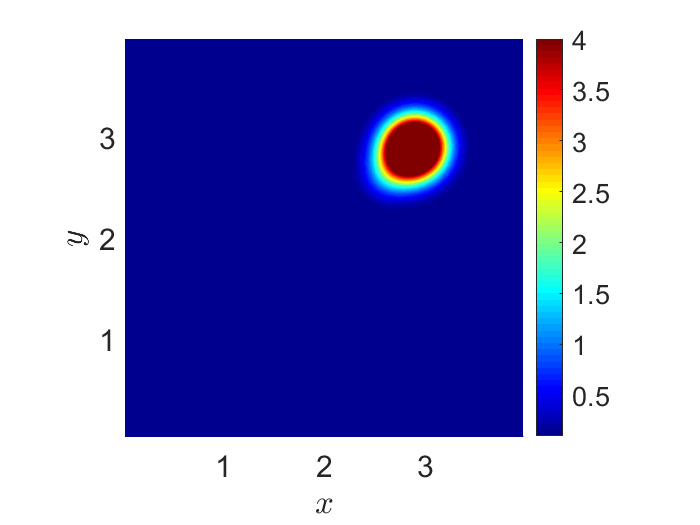}}
\\
\subfigure[$t=5$]{\includegraphics[scale=0.25]{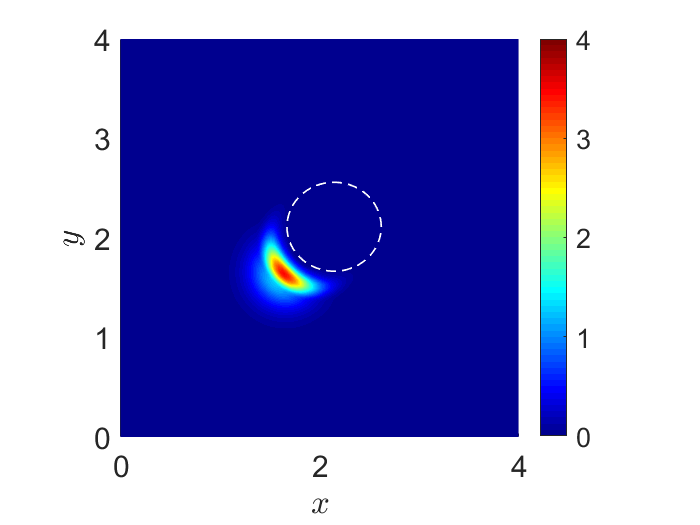}}
\subfigure[$t=10$]{\includegraphics[scale=0.25]{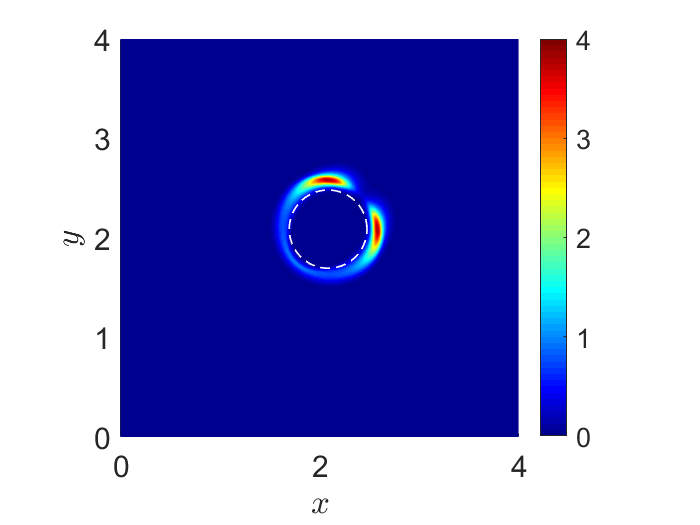}}
\subfigure[$t=50$]{\includegraphics[scale=0.25]{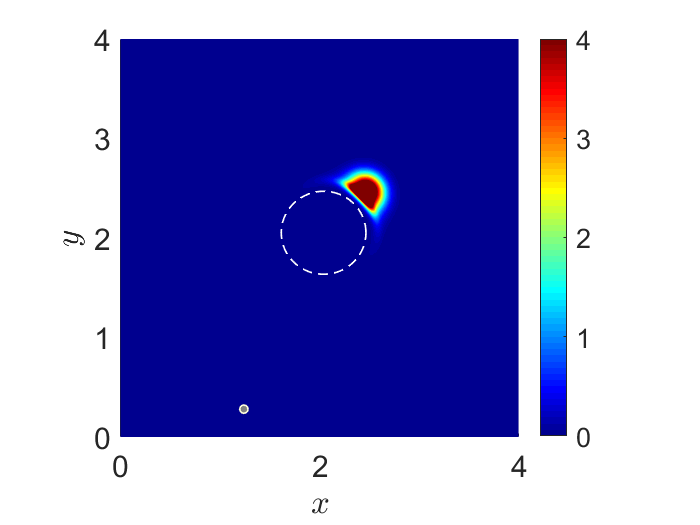}}
\\
\subfigure[$t=5$]{\includegraphics[scale=0.25]{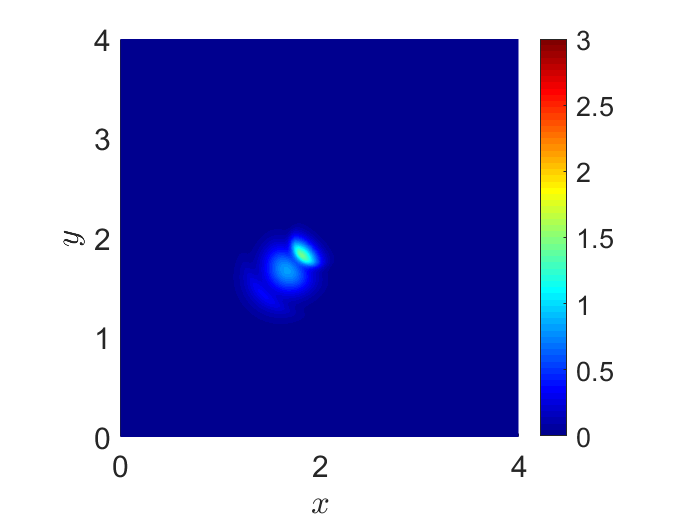}}
\subfigure[$t=10$]{\includegraphics[scale=0.25]{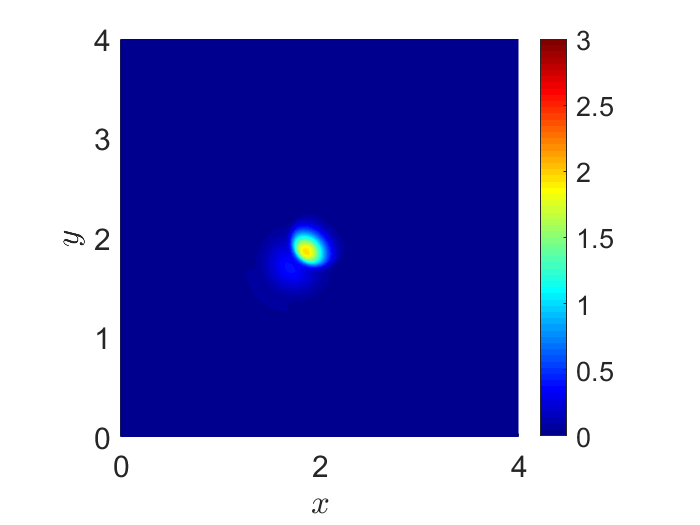}}
\subfigure[$t=50$]{\includegraphics[scale=0.25]{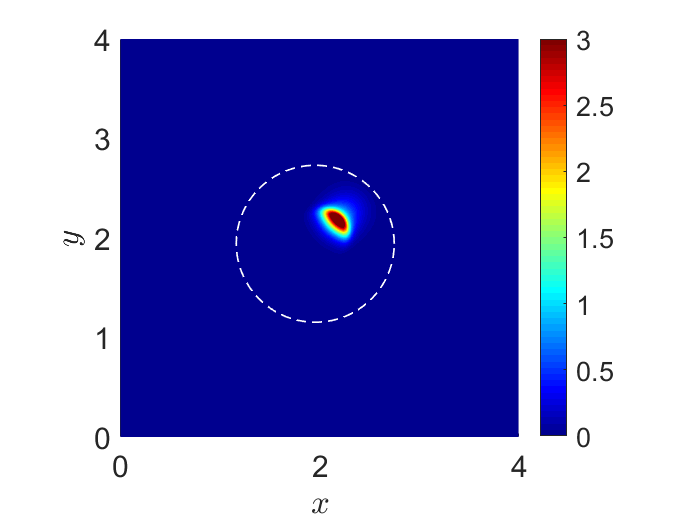}}
\end{center}
\caption{Motion of a cell cluster initially distributed as in (a) toward a chemoattractant with distribution as in (b) passing through a dense ECM region as in (c) (In the plotted case $M_b=0.2$ and $M_m=1.2$).  
The sensing kernels are Heaviside functions and the sensing radii are $R_{\scS}^{max}=R_{\scriptscriptstyle M} ^{\max}=0.3$. The turning frequency is $\mu=10$. 
(d-f) $M_m=1.2$ is below $M_{th}=1.39$, $M_b=0.2$ above $M_{0}=0.01$, and $M_{max}=0.7$. 
(g-i) $M_m=1.2$ is above $M_{th}=0.8$, $M_b=0.5$ above $M_{0}=0.2$, and $M_{max}=0.5$. The region within the white dashed circle is then too dense because $M>M_{th}=0.8$. 
(j-l) $M_m=0.7$ is below $M_{th}=1.5$, $M_b=0$ is below $M_{0}=0.1$, and $M_{max}=0.8$. In figure (l), 
the region outside the white dashed circle has poor ECM with $M<M_0=0.1$. }
\label{ecm_mix}
\end{figure}

In this section we shall consider the motion of a cell cluster toward a chemoattractant in a strongly heterogeneous environment. Specifically, the initial distribution of cells is
$$\rho_0(\x)=0.5\exp\left[-\dfrac{|\x-\x_0|^2}{0.2}\right]\quad {\rm with} \quad \x_0=(1.5, 1.5),$$ 
as in Fig. \ref{ecm_mix}(a), that of the chemoattractant is 
$$C(\x)=50\exp\left[-\dfrac{|\x-\x_c|^2}{0.6}\right]\quad {\rm with} \quad \x_c=(3, 3),$$ 
as in Fig. \ref{ecm_mix}(b), and that of the ECM is
$$M(\x)=M_b+(M_m-M_b)\exp\left[-\dfrac{|\x-\x_M|^2}{0.4}\right]\quad {\rm with} \quad \x_M=(2, 2),$$
as in Fig. \ref{ecm_mix}(c).
So, in a homogeneous environment cells would tend to move more or less along the diagonal of the square domain toward the maximum concentration of chemoattractant, while in the heterogeneous case they would tend to avoid the region with too dense ECM.

In the first simulation reported in the second row of Fig. \ref{ecm_mix} the ECM density is distributed in a way that it does not represent physical limits of migration. In fact, its maximum concentration $M_m=1.2$ is below the threshold value $M_{th}=1.39$ and its minimum concentration $M_b=0.2$ is above the minimum value of ECM density $M_0=0.01$ for crawling. So, cells are not blocked, but they are however slowed down as the maximum speed is achieved where $M=M_{max}=0.7$. This means that they tend to move faster around the peak of  concentration of ECM, with some trapping for a longer period of those cells that move diagonally. Having passed the denser ECM region, cells cluster again and move towards the chemoattractant (see also Supplementary Movie $ECM1.mp4$).

In the second simulation reported in the third row of Fig. \ref{ecm_mix} the maximum concentration of ECM $M_m=1.2$ is above the threshold value $M_{th}=0.8$. So, cells can not enter the region that is identified by the white dashed circle in Fig. \ref{ecm_mix}(g)-(i), moving around it to join again at the north-east pole of the denser region. Then, the clusters merge again and move toward the maximum of chemoattractant.
(see also Supplementary Movie $ECM2.mp4$).


In the third simulation reported in the bottom row of Fig. \ref{ecm_mix} as for the simulation reported in the second row, ECM density is not a prohibitive obstacle, being the maximum concentration of the ECM $M_m=0.7$ that is lower than the threshold value $M_{th}=0.8$. On the other hand, the value $M_0=0.1$ is such that outside the dashed circle cells do not have not enough ECM to anchor. In this situation, initially cells oriented along the diagonal are able to grab the denser region of ECM because of a sensing radius $R_{M} ^{\max}=0.3$, while motion in other directions is hampered if not completely inhibited. Cells then cluster in the region with a comfortable density of ECM towards the north-east side of the circle, because they are attracted towards the maximum chemical concentration. However, ahead they sense a region lacking of ECM for adhesion and remain stuck.   
(see also Supplementary Movie $ECM3.mp4$).

%
%
%

\section{Discussion}

The kinetic model developed in this article is based on the observation that 
\begin{enumerate}[label=\roman*)]
\item cells sense their environment collecting chemical and mechanical cues  by extending protrusions that can be much longer than the cell diameter; 
\item the information acquired determine cell polarization and speed;
\item the mechanisms governing cell polarization and speed depend on different intracellular mechanisms;
\item the valuable information can be limited by the presence of physical limits of migration, such as cell overcrowding,  cell layers, like mesothelial or endothelial linings, basal membranes, or in general ECM with pores too small to be penetrated by the cell nucleus or even by its protrusions. 
\end{enumerate}
From the modelling point of view, these points respectively imply that the kinetic model is characterized by (i) non-local turning operators ($\ie$, the integrals over $\lambda$ and $\lambda'$), (ii) a probability distribution that depends on a speed $v$ and an orientation unit vector $\hv$, (iii)  with a turning operator split in a part influencing $v$ and a part influencing $\hv$ depending on different sensing kernels ($\gamma_{\scS}$ and $\gamma_{\scS'})$ and cues ($\cS$ and $\cS'$) (iv) operating on domains (identified by $R_{\scS}^{\scM}$ and $R_{\scS'}^{\scM'}$) that can depend on the presence of physical limits of migration, such as those mentioned above. 

The article shows in particular how  important it is to handle and model properly situations that might look extreme but actually characterize many physiological and pathological situations leading to cell aggregation and collective migration, to cell compartmentalization by basal membranes, to cell invasion when the membranes rupture or cells acquire a phenotype that allows them to pass through their narrow pores, leading to intravasation and extravasation of metastasis.  

The model is very flexible and was applied to situations taking into account of volume filling effects and cell-cell adhesion.
Cell-matrix interaction was also considered both in the case of thick ECM and when a lack of ECM might hamper the formation of focal adhesions that are essential for cell crawling. In the latter case it was shown that if the cell is able to extend its protrusion beyond the problematic area to reach a region where focal adhesions can be formed again, then, due to the non-local character of the model, it is able to cross over the region with poor ECM. Otherwise, it is segregated close to the border of the area lacking of ECM. A virtual experiment of motion along stripes of laminin is also performed.   

Simulations show how pattern may form spontaneously from nearly homogeneous configurations, especially when the sensing kernel is a Dirac delta. The characteristic size of the pattern is related to the sensing radius. The presence of this effect calls for a stability analysis that will be performed in a future article.

Another important topic to be addressed that is not included in the present model is how to handle different cues  governing either sensing or speed. The former aspect refers, for instance, to cases in which cells adhere to each other while under the action of chemotaxis and/or haptotaxis. Being able to deal con temporarily with the two phenomena, together with volume filling, is fundamental to deal with collective chemotaxis.
The latter aspect refers, for instance, to overcrowding due to strongly heterogeneous distributions of ECM and, in particular, volume filling in presence of basal membranes.

\section*{Acknowledgements}
This work was partially supported by Istituto Nazionale di Alta Matematica, Ministry of Education, Universities and Research, through the MIUR grant Dipartimenti di Eccellenza 2018-2022 and Compagnia di San Paolo that finances NL's Ph.D. scholarship.

\bibliography{references}
\end{document}